\documentclass[preprint,twoside,english,sort&compress]{elsarticle}
\usepackage[left=30mm,width=150mm,top=25mm,bottom=25mm]{geometry}
\usepackage[T1]{fontenc}
\usepackage[latin9]{inputenc}
\usepackage{amstext}
\makeatletter
\journal{}
\usepackage{xcolor}
\usepackage{amsmath}
\usepackage{amssymb}
\usepackage{slashbox,pict2e} 

\usepackage{ctex}

\usepackage{Sweave}

\makeatother
\usepackage{amsmath,bm}
\usepackage{algorithm}
\usepackage{algorithmic}
\usepackage{babel}
\def\varneg{\mathord{\raise1.41ex\hbox{\vrule width.4em height.4pt}\kern0pt\vrule width.4pt height1.5ex}}

\graphicspath{{./fig/}}
\begin{document}

\begin{frontmatter}

\title{A similarity-based community detection method with multiple prototype representation}

\author[rvt,focal]{Kuang Zhou\corref{cor1}}

\ead{kzhoumath@163.com}

\author[focal]{Arnaud Martin}


\author[rvt]{Quan Pan}
\cortext[cor1]{Corresponding author}
\address[rvt]{School of Automation, Northwestern Polytechnical University, Xi'an, Shaanxi 710072, PR China}
\address[focal]{DRUID, IRISA, University of Rennes 1, Rue E. Branly, 22300 Lannion, France}
\begin{abstract}
Communities are of great importance for understanding graph structures in social networks.  Some existing community detection algorithms use a single prototype to represent each group. In real applications, this may not adequately model the different types of communities  and hence limits the clustering performance on social networks. To address this problem, a Similarity-based Multi-Prototype (SMP) community detection approach is proposed in this paper. In SMP, vertices in each community carry various weights  to describe their degree of representativeness. This mechanism enables each community to be represented by more than one node. The centrality of nodes is used to calculate  prototype weights, while similarity is utilized to guide us to partitioning the graph.  Experimental results on computer generated and real-world networks clearly show that SMP performs well for detecting communities. Moreover, the method could provide richer information for the inner structure of the detected communities with the help of  prototype weights compared with the existing community detection models.
\end{abstract}


\end{frontmatter}

\section{Introduction}
In order to have a better understanding of organizations and functions in real-world networked systems, the community structure in the graph is a primary feature that should be taken into consideration \citep{fortunato2010community}. As a result,
community detection, which can extract specific structures from
complex networks,  has attracted considerable attention crossing many areas from physics, biology, and economics to sociology \citep{costa2011analyzing,girvan2002community}, where systems are often represented as graphs.
Generally, a community in a network is a subgraph whose nodes are densely connected within itself but sparsely connected with the rest of the network \citep{newman2006modularity,zhang2007identification,jiang2012efficient}.

Recently, significant progress has been achieved in this research field and several popular algorithms for community detection have been presented.
One of the most popular type of classical methods partitions networks by optimizing some criteria.  \citet{newman2004finding} proposed a network modularity measure (usually denoted by $Q$) and several algorithms that try to maximize $Q$ have been
designed \citep{blondel2008fast,clauset2004finding,duch2005community}.  But recent researches have found that the modularity based algorithms could not detect communities smaller than a certain size.
This problem is famously known as the resolution limit \citep{fortunato2007resolution}.  The single optimization criteria, {\em i.e.,} modularity, may not be adequate to represent the structures in complex networks, thus \citet{amiri2013community} suggested a new  community detection process as a multi-objective optimization problem.
Another family of approaches considers hierarchical clustering techniques. It merges or splits clusters according to a topological measure of similarity between the nodes and tries to  build a hierarchical tree of partitions \citep{lancichinetti2009detecting,huang2011density,yang2013hierarchical,kim2013spanning,zhang2014mining,kim2015detecting}.
Also there are some ways, such as spectral methods
\citep{smyth2005spectral} and signal process method
\citep{hu2008community,jiang2012efficient}, to map topological
relationship of nodes in the graphs into geometrical structures of vectors in
$n$-dimensional Euclidean space, where  classical clustering methods like
classical $C$-Means (CM) \citep{jiang2012efficient}, Fuzzy $C$-Means (FCM) \citep{zhang2007identification, hu2008community} or Evidential $C$-Means (ECM) \citep{zhou2014evidential} could be evoked. However, there must be some loss of information during
the mapping process. Besides, these prototype-based partition  methods  themselves are sensitive to the initial seeds. For social networks with good  community structures, the center of one group is likely to be one person, who plays the leader role in the community. That is to say, one of the  members in the group is better to be selected as the seed, rather than the center of all the objects.

\begin{center} \begin{figure}[!thbt] \centering
		\includegraphics[width=0.45\linewidth]{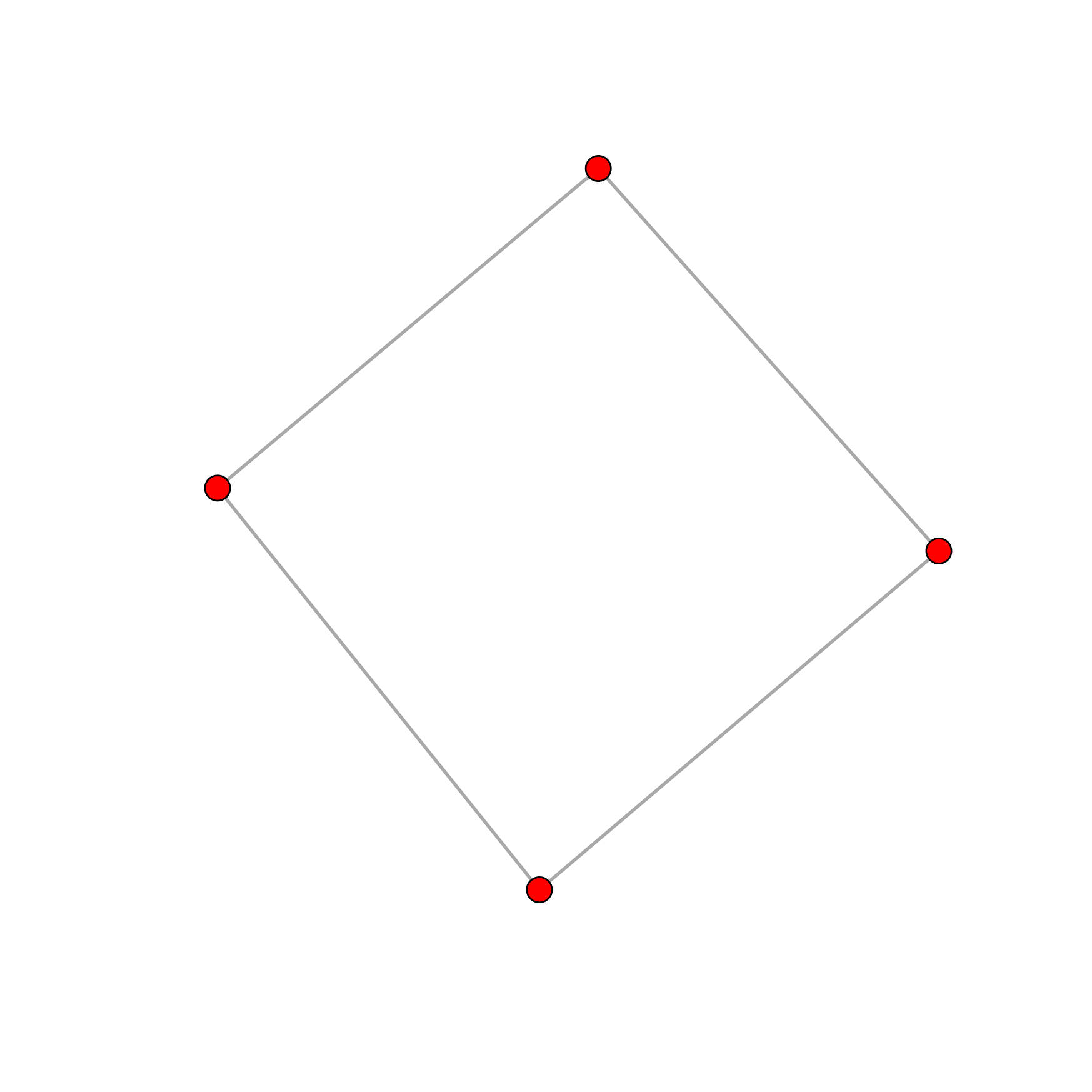}
		\hfill
		\includegraphics[width=0.45\linewidth]{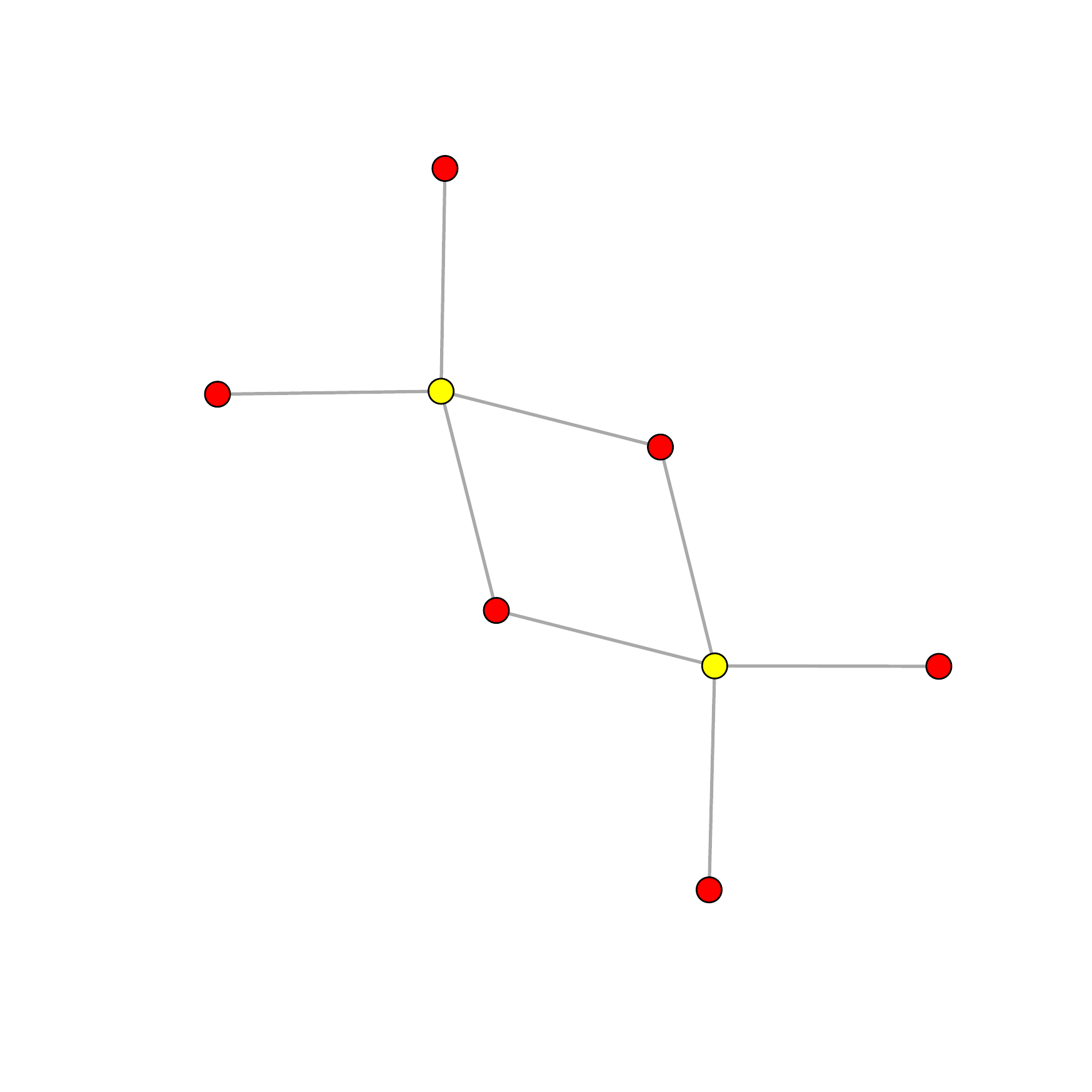}
		\hfill \parbox{.45\linewidth}{\centering\small a. Community 1}
		\hfill \parbox{.45\linewidth}{\centering\small b. Community 2}
 \hfill \caption{Two small community's structures. (For interpretation of the references to color in this figure legend, the reader is referred to the web version of this article.)} \label{simplenetwork} \end{figure} \end{center}

To solve these problems, \citet{jiang2012efficient} proposed an efficient algorithm named $K$-rank which selects the node with the highest centrality value as the prototype. In our previous work, an evidential centrality measure is used to set one ``most possible" object in the class to be the prototype \citep{zhou2015median}. We believe that the characteristic on the prototype of each community is important for community detection. However, in some cases the way of using only one node to describe a community may not be sufficient enough. To illustrate the limitation of one-prototype community representation, we use two simple community structures shown in Figure~\ref{simplenetwork}.  The first community consists of four members while the second has eight. It can be seen that in the left community, it is unreasonable to describe the cluster structure using any one of the four nodes in the group, since no one of the four nodes could be viewed as a more proper representative than the other three. In the right community in Figure~\ref{simplenetwork}, two members (marked yellow) out of the eight are equal reasonable to be selected as the representative of the community. This means choosing any one of them may fail to detect the complete set of all the candidate representative nodes. From these examples, we can see that for some networks, in order to capture various aspects of the community structures, we may need more members rather than one to be referred as the prototypes of an individual group.

Motivated by this idea, in this paper, a Similarity-based Multiple Prototype (SMP) community detection approach  is proposed. The centrality values are used as the  criterion to select multiple prototypes  to characterize each community, and the prototype weights are derived to describe the degree of representativeness of the related objects for their own community. Then the similarity between each node and community is defined, and the nodes are partitioned into divided communities according to these  similarities. 
 Here, we emphasize some key points
different from those earlier studies and the contribution of this work. Firstly, although there are some multi-prototype clustering methods for the classical data sets \cite{liu2009multi,wang2014incremental}, there is little such work for community detection problems. Here a new community representation mechanism using multiple prototypes is proposed. Experimental results on artificial and real-world networks show that multiple prototypes are more powerful than a single center for representing a community,  especially for the graphs without clear community structures.  Secondly,  the concept of prototype weights is presented, which describes the degree of representativeness
of a member in its own group. With the help of prototype weights, SMP provides more sufficient description for each individual community.   This enables us to gain a deep insight into the internal structure of a community, which we
believe is also very important and useful for network analysis.
Thirdly, in the proposed community detection approach, different kinds of similarity and centrality measures could be adopted, which makes it more practical and flexible in real applications.

The rest of this paper is organized as follows. In Section 2, some basic concepts and the rationale of our method are briefly introduced. In Section
3, the multi-prototype community detection approach is presented in detail. In order to show the effectiveness
of our approach, in Section 4 we test our algorithm on different artificial and real-world networks and make comparisons with the existing methods. Finally, we conclude and present some perspectives in Section 5.

\section{Preliminary knowledge}
In this section some background knowledge related to community detection problems and social networks, including centrality and similarity measures, modularity and some classical existing algorithms, will be presented.
\subsection{Node centrality and similarity}
Generally speaking, the person who is the center of a community in a social
network has the following characteristics: he has relation with most of  the
members of the group and the relationships are stronger than usual; he  may
directly contact with other persons who also play an important role in their
own communities.  Therefore, the centers
of the community should be set to the ones not only with high degree and weight strength,
but also with neighbors  who also have high degree and strength. The degree of node is the number of its connections with other nodes, and the strength describes the levels of these connections.  \citet{gao2013modified} proposed an evidential centrality measure, named Evidential Semi-local Centrality (ESC), based on the theory of belief functions. In the application of ESC, the degree and strength of each node are first expressed by basic belief assignments (BBA), and then the fused importance is calculated using the combination rule in the theory of belief functions. The higher the ESC value is, the more important the node is. \citet{gao2013modified} pointed out that it is
more efficient than the existing centrality measures
such as Degree Centrality (DC), Betweenness Centrality
(BC) and Closeness Centrality (CC). The detail computation process of ESC can be found in  \citep{gao2013modified}.

The similarity measures the closeness between any pair of nodes in the graph. In \citep{zhou2009predicting} several node similarity metrics on basis of local information were described and the performance of different measures applied to community detection was discussed. Here we give a brief description of some measures. Let $G(V,E)$ be an undirected network, where $V$ is the set of  $N$ nodes and $E$ is the sets of $m$ edges. Let $\bm{A}=(a_{ij})_{N\times N}$ denote the adjacency matrix, where $a_{ij}=1$ represents that there is an edge between nodes $i$ and $j$.

\begin{description}
  \item [(1)] Common neighbors. This measure is based on the idea that more common neighbors the pair shares, more similar they are. Thus the similarity can be simply  proportional to the number of their shared neighbors:
    \begin{equation}
    \label{sim_com_eq}
     s^{\text{C}}(x,y)=|N(x)\cap N(y)|,
\end{equation}
where $N(x)=\{w\in V\setminus x: a(w,x)=1\}$ denotes the set of vertices that are adjacent to $x$.

  \item [(2)] Jaccard Index. This index was proposed by Jaccard over a hundred years ago, and is defined as
   \begin{equation}\label{Jaccard}
            s^{\text{J}}(x,y)=\frac{|N(x) \cap N(y)|}{|N(x) \cup N(y)|}.
        \end{equation}
  \item [(3)] Zhou-L\"u-Zhang Index. \citet{zhou2009predicting} also proposed a new similarity metric which is motivated by the resource allocation process:
       \begin{equation}
         s^{\text{Z}}(x,y)=\sum_{z \in N(x) \cap N(y)} \frac{1}{d(z)},
       \end{equation}
where $d(z)$ is the degree of node $z$.
\end{description}

\citet{pan2010detecting} pointed out that the similarity measure proposed by \citet{zhou2009predicting} may bring about inaccurate results for community detection on the networks as the metric can not differentiate the
tightness relation between a pair of nodes whether they are connected directly or indirectly. In order to overcome this defect, in his presented new measure the similarity between unconnected pair is simply set to be 0:
\begin{equation}
  S^P(x,y)=\begin{cases}
   \sum\limits_{z \in N(x) \cap N(y)} \frac{1}{d(z)}, & \text{if}~ x, y ~\text{are connected},\\
   0 & \text{otherwise}.
  \end{cases}
\end{equation}

A similarity measure considering the global graph structure is put forward by \citet{hu2008community} based on signaling propagation in the network.  For a network with $N$ nodes, every node is viewed as an excitable system which can send, receive, and record signals. Initially, a node is selected as the source of signal. Then
the source node sends a signal to its neighbors and itself first.
Afterwards, the nodes with signals can also send signals to
their neighbors and themselves.  After a
certain $T$ time steps, the amount distribution of signals over
the nodes could be viewed as the influence of the source
node on the whole network. Naturally, compared with nodes in other communities, the nodes of the same community have more similar
influence on the whole network. Therefore, similarities between nodes could be obtained by calculating the differences between the amount of signals they have received.

\subsection{Modularity}
Recently, many criteria were proposed for evaluating the partition of a network. A widely used measure
called modularity, or $Q$ function was presented by \citet{newman2004finding}.
Let $G(V,E,W)$ be an undirected network, $V$ is the set of  $N$ nodes, $E$ is the set of edges, and $\bm{W}$ is a $N \times N$ edge weight matrix with elements $w_{ij}, i,j=1,2,\cdots,N$. Given a hard
partition with $K$ groups $\bm{U}=(u_{ik})_{N
\times K}$, where $u_{ik}$ is one if vertex $i$ ($i=1, 2, \cdots, N$) belongs to the $k_{th}$ ($k=1, 2, \cdots, K$)
community, 0 otherwise. Denote the $K$ crisp subsets of vertices by
$\{C_1,C_2,\cdots,C_K\}$,  then the modularity can be defined as \citep{fortunato2010community}:
 \begin{equation}\label{hard_modu2}
	Q_h=\frac{1}{\left \| \bm{W} \right \|}\sum_{k=1}^K \sum_{i,j \in C_k}
	\left(w_{ij}-\frac{k_i k_j}{\left \| \bm{W} \right \|}\right), \end{equation}
where $\left \| \bm{W} \right \|=\sum_{i,j=1}^N w_{ij}$, $k_i=\sum_{j=1}^N w_{ij}$.

The $Q$ measure has been proved highly effective in practice for community evaluation, although  \citet{fortunato2007resolution} claim resolution limits of modularity-based division methods. Besides, some other problems of Newman's modularity have also been found \citep{chen2009detecting}. To solve these problems, some new modularity measures have been proposed \citep{scripps2007exploration,chen2009detecting}. In this paper, the Max--Min (MM) modularity function proposed  by \citet{chen2009detecting} is utilized as the index to determine the optimal number of communities. MM modularity attempts to maximize the number of edges within groups and minimize the number of unrelated pairs from the user-defined unrelated pair set within groups at the same
time:
\begin{equation}
  Q_{\text{MM}}=Q_{\text{max}}-Q_{\text{min}},
  \label{MM}
\end{equation}
where $Q_\text{max}$ is the $Q$ modularity of the original graph, while $Q_\text{min}$ is that of the complement graph $G^{'}$.  Graph $G^{'}=(Y,E^{'})$ is created based on the user-defined criteria $\mathcal{M}$ which defines  whether two disconnected nodes $i,j$ are related $(i,j)\in \mathcal{M}$ or unrelated $(i,j)\notin \mathcal{M}$, {\em i.e.,} $(i,j) \in E^{'}$ if $(i,j) \notin E$ and $(i,j)\notin \mathcal{M}$. The related pairs $\mathcal{M}$ can be given by experts, or defined according to the original structure \citep{chen2009detecting}.

\subsection{Some classical methods of community detection}
\label{classicalmethods}
In Section \ref{test} we will compare the proposed algorithm with five existing methods:  $K$-rank algorithm \citep{jiang2012efficient}, Multi-level Modularity Optimization (MMO) algorithm \citep{blondel2008fast}, Leading Eigenvector (LE) algorithm \citep{newman2006finding}, Label Propagation (LP)  algorithm \citep{raghavan2007near}, and Information Map (InfoMap) algorithm \citep{Rosvall29012008}. Thus here we give a short presentation of these   five approaches.

\textbf{MMO} is a heuristic method  based on modularity optimization, and the algorithm is divided into two phases repeated iteratively. In the beginning of the first phase, the network is thought to have $N$ groups each of which consists of only one node. Then for each node $i$, it may be placed into a new community (it must be a community that one of its neighbors belongs to) for which the  gain of modularity is maximum. The first phase is not completed until no further improvement of the modularity can be achieved. The second phase consists in building a new network whose nodes are the communities detected in the last phase, and then the first phase can be reapplied on this newly created graph. \citet{blondel2008fast} pointed out that MMO outperformed all other known community detection methods in terms of computation time.

\citet{newman2006finding} demonstrated that the modularity can be succinctly expressed as a function of the eigenvalues and eigenvectors of the modularity matrix and derived a competitive Leading Eigenvector (\textbf{LE}) algorithm  for identifying communities. The graph is first divided  into two groups according to the signs of the
elements of the eigenvector corresponding to the most
positive eigenvalue of the modularity matrix, and then can be partitioned into more communities depending on the requirement analogously. It is showed that LE works better than the standard spectral partitioning method as it is unconstrained by the need to find groups of any particular size \citep{newman2006finding} .

\textbf{LP} is investigated by \citet{raghavan2007near} and it only uses the network structure  and requires neither optimization of a predefined objective function nor prior information about the communities. In this model every node is initialized with a unique label. Afterwards each node adopts the label that most of its neighbors currently
have at every step. In this iterative process densely connected groups of nodes form a consensus on a unique label to form communities.

\textbf{InfoMap} uses the probability flow of random walks on a network as a proxy for information flows in the real system, and
graph clustering turns then into the coding problem of finding the partition
that yields the minimum description length of an infinite random walk~\citep{fortunato2010community}.  The network is optimally decomposed into modules by compressing the information needed to describe of the
process of information diffusion across the graph \citep{Rosvall29012008}. The regularities in the community structure and their relationships are reflected by a map.

$\bm{K}$\textbf{-rank} algorithm is proposed by \citet{jiang2012efficient}, and it uses an alternate iteration  strategy like $K$-means. Firstly, the top--$K$ nodes with the highest rank centrality is selected as initial seeds. This initialization mechanism could overcome the problem brought by the random initial centers in the application of  prototype-based clustering methods like $K$-means. Then the seeds and cluster labels are updated alternately by using an iterative technique. As illustrated before, the way of selecting $K$ representative members with each to totally represent one individual community may be insufficient to fully characterize a community. This in turn indicates that multiple nodes should be utilized in order to capture each group in the network more accurately. 
\section{The multi-prototype community detection approach}
\label{our_method}
We propose here our method. After an introduction of  the concept of representative weights (also called prototype weights) in Section \ref{section_prowei},  the whole algorithm will be presented in detail in Section 3.2. The problem of determining the optimum community  number and the complexity of the algorithm will be discussed in Section 3.3 and 3.4 respectively.
\subsection{The prototype weights}
\label{section_prowei}
Suppose $C=\{C_1, C_2, \cdots, C_K\}$ is a partition of a graph $G(V, E)$, where $V$ is the set of nodes and $E$ is the set of edges.  The $N$ nodes in the graph can be denoted by $\{n_1, n_2, \cdots, n_N\}$. The matrix $\bm{V}_{K\times N}$ denotes the prototype weights of $N$ nodes with respect to all the $K$ communities. As analyzed before, the centrality value of a node can be used to express the belief that the node plays the center role in its community. Therefore, the probabilistic weight of node $j$'s degree of representativeness in cluster $C_r$ can be derived as below:
 \begin{equation}
 \label{prowei}
   V_{rj}=\begin{cases}
     \frac{P_r(j)}{\sum\limits_{\{h: n_h\in C_r\}} P_r(h)} & n_j \in C_r \\
     0 & n_j \notin C_r,
   \end{cases} ~~~ r=1,2,\cdots,K, ~j=1,2,\cdots N,
 \end{equation}
where $P_r(j)$ is the centrality of node $n_j$ in the subgraph corresponding community $C_r$. Then, for a given node $n_i$, the similarity between $n_i$ and community $C_j$, denoted by $\bar{s}_{ij}$,  can be obtained as
\begin{equation}
\label{dij}
  \bar{s}_{ij}=\sum_{h=1}^{N}v_{jh}s_{ih},
\end{equation}
where $s_{ih}$  is the similarity between nodes $n_i$ and $n_h$. From Eqs.~\eqref{prowei} and \eqref{dij} we can see that $\bar{s}_{ij}$ is a weighted sum of the similarity between node $n_i$ and all the nodes in community $C_j$, and  the weights used in the  summation  depend on the contribution of the nodes to their own community.
\label{initpro}
\subsection{The detection algorithm}
The whole SMP  algorithm  to detect communities in social networks is summarized as Algorithm \ref{alg:method}. In fact SMP is a variation of $K$-means, $K$-medoids and $K$-rank. The difference between SMP and the other three clustering algorithms lies in  the manner of updating the prototypes. $K$-means uses the average value to represent every class while $K$-medoids and $K$-rank uses one ``most possible" object.   On the contrary, SMP adopts an effective multi-prototype representation  based on the determined prototype weights of each member in the group.  Due to the various types of community structures, the way to represent a cluster using multiple prototypes is more reasonable in real applications.  Moreover, SMP often needs fewer iterations than $K$-means to make the algorithm convergent.

\begin{algorithm}\caption{\textbf{:} ~The Similarity-based Multi-Prototype (SMP) community detection algorithm}\label{alg:method}
\begin{algorithmic}
\STATE {\textbf{Input:} $K$, the number of communities;  $\bm{A}$, the adjacency matrix; $\bm{W}$, the weight matrix (if any);  $N_\text{max}$, the maximum number of iterations.}
\STATE {\textbf{Initialization:}\\ (1). Select the top $K$ nodes  with highest centralities as the initial $K$ prototypes. \\(2). Calculate the similarity matrix between any two nodes in the graph. \\(3). Extract the similarity matrix between the nodes and the prototypes. Partition the node into the community to which its nearest prototype belongs, and get the initial $K$ classes of the graph: $C_1, C_2, \cdots, C_K$.}
\REPEAT
\STATE{
 (4). Update the matrices $\bm{V}_{K\times N}$ recording  prototype weights of $N$ nodes with respect to all the $K$ communities based on the current partitions using Eq.~\eqref{prowei}. \\
 (5). Calculate the similarity between node $n_i$ and community $C_j$, $\bar{s}_{ij}$, using Eq.~\eqref{dij}, and then cluster the vertices into $k$ communities with every node being in the community it is most similar to.
}
\UNTIL{All the detected communities remain unchanged or the number of iterations comes to $N_\text{max}$.}
\STATE {\textbf{Output:} The membership of each node and  the prototype weights of all the members in each community.}
\end{algorithmic}
\end{algorithm}

\textbf{Remark:} As we can see, SMP provides us a crisp (hard) partition of the analyzed network. Also the similarity between node $n_i$ and community $C_j$ could be obtained by Eq.~\eqref{dij}. Then the node $n_i$'s membership  with regard to community $C_j$ can be defined as follows:
\begin{equation}
\label{fuzzymemb}
  u_{ij}=\frac{\bar{s}_{ij}}{\sum\limits_{h=1}^{K}\bar{s}_{ih}}, ~~~i=1,2,\cdots,N, ~j=1,2,\cdots,K.
\end{equation}
This form of membership measure is in line with that got by FCM algorithm,  where the  membership values  assigned
to an object are inversely related to the relative distance to
the cluster. Similarly here the memberships in Eq.~\eqref{fuzzymemb} are determined by the relative similarities. One of the problem of fuzzy membership has been reported is that it could not  distinguish between ``equal evidence" (membership values
are large and equal for a number of alternatives) and ``ignorance" (all the membership values are equal but very close to zero) \citep{krishnapuram1993possibilistic,pal2005possibilistic}.  If node $n_i$ is equidistant from more than one community, the membership of each cluster will be the same,
regardless of the absolute values of the similarity to the communities. Consequently, the fuzzy membership could not be applied to detect noise objects (outliers) which are far but equidistant to some  communities~\citep{pal2005possibilistic}. In SMP, the prototype weights can help us solve this problem, which we will show in detail in Section \ref{realdata}.
\subsection{Determining the number of communities}
In the first step of SMP algorithm, the additional information about the number of  communities ($K$) should be specified. This is  also a fundamental issue in classical CM and FCM clusterings. In fact, to determine the optimal number of clusters is an open problem for prototype-based clustering methods. Most of the methods to solve this problem consist in computing a validity index from several community structures detected with different values of $K$ and looking for a minimum or maximum of a given criterion \citep{hu2008community,zhang2007identification,nepusz2008fuzzy}. In this paper MM-modularity (Eq.~\eqref{MM}) is used to estimate a proper $K$. The modularity values signify the quality of the detected communities. When the modularity achieves the maximum, we can get the best $K$.
\subsection{The complexity of SMP algorithm}
The complexity of SMP consists of calculating similarities and centralities of nodes and iterative process. If we use signal similarity and evidential semi-local centrality measures, as we will see in Section 4, the corresponding time complexity if $\text{O}(c(|k|+1)N^2)$ \citep{hu2008community} and $\text{O}(N|k|^2)$ \citep{gao2013modified}, where $c$ is the number of propagation, $|k|$ is the average degree of vertices in the network, and $N$ is the number of nodes. The iterative technique  is similar to that in $K$-means. The only difference is the strategy of updating the prototypes. $K$-means computes the average value of all the members in the cluster, while SMP tries to find prototype weights of all the members. As the communities are subgraphs which are much smaller than the original network, the updating prototype weights process of SMP does not cost much. If the number of communities $K$ is fixed, the time complexity of $K$-means clustering is $\text{O}(NKt)$, where $t$ is the number of iterations. Consequently, the total complexity of SMP is $\text{O}(c(|k|+1)N^2+N|k|^2+NKt)$. It is worth noting that SMP often needs fewer iterations.
\section{Experimental results}
\label{test}
In this section some experiments are performed on both
computer-generated graphs and real-world networks whose community structure is known in advance. Apart from $K$-rank \citep{jiang2012efficient},
we also compare SMP  with four other classical methods:  Multi-level Modularity Optimization (MMO) algorithm \citep{blondel2008fast}, Leading Eigenvector (LE) algorithm \citep{newman2006finding}, Label Propagation (LP) algorithm  \citep{raghavan2007near}, and Information Map algorithm (InfoMap)   \citep{Rosvall29012008} presented in Section \ref{classicalmethods}. The obtained
community structures are evaluated with known performance measures, i.e., accuracy and NMI (Normalized
Mutual Information). As the benchmarks and the real-world data sets used in this paper are with known community structure,  accuracy and NMI measure the similarity between the planted partitions (ground truth)
and the results of the algorithms. The NMI of two partitions $A$ and $B$  of the graph, $I(A,B)$, can be calculated  by
\begin{equation}
\label{nmi_eq}
 I(A,B)=\frac{-2\sum_{i=1}^{C_A}\sum_{j=1}^{C_B}N_{ij}\log (\frac{N_{ij}n}{N_{i\cdot}N_{\cdot j}})}{
  \sum_{i=1}^{C_A}N_{i\cdot} \log(\frac{N_{i\cdot}}{n})+\sum_{j=1}^{C_B}N_{\cdot j} \log(\frac{N_{\cdot j}}{n})},
\end{equation}
where $C_A$ and $C_B$ denote the numbers of communities in partitions $A$ and $B$ respectively. The notation $N_{ij}$ denotes the element of matrix $(\bm{N})_{C_A \times C_B}$, representing the number of nodes in the $i_{th}$ community of $A$ that appear in the $j_{th}$ community of $B$. The sum over row $i$ of matrix $\bm{N}$  is denoted by $N_{i\cdot}$ and that over column $j$ by $N_{\cdot j}$.
Both accuracy and NMI measure the proportion of the nodes that have been grouped correctly, and represent the consistence between the found  community structure and the presumed one \citep{fan2007accuracy,hu2008community}. 
The influence of different similarity and centrality measures in the application of SMP will be discussed in the first experiment. After that  we will use the evidential semi-local centrality and signal similarity in the following tests based on the experimental results.

\subsection{Computer-generated graphs}

The algorithm is  first compared by means of two  classes of  computer-generated artificial benchmark networks, namely, Girvan and Newman \citep{girvan2002community} (GN) and
\citet{lancichinetti2008benchmark} benchmark (LFR) networks. For the former, each network has $N=128$ nodes in total and 32 nodes in each of the four divided communities. The average degree of each vertex is set to 16. For a given node, the average number of links to its fellows in the inner community, denoted by $Z_\text{in}$, is varied from 8 to 16. The average number of edges between communities, denoted by $Z_\text{out}$, is varied from 8 to 0. The larger $Z_\text{in}$ is, the more apparent community structure the network has.

\begin{center} \begin{figure}[!thbt] \centering
		\includegraphics[width=0.45\linewidth]{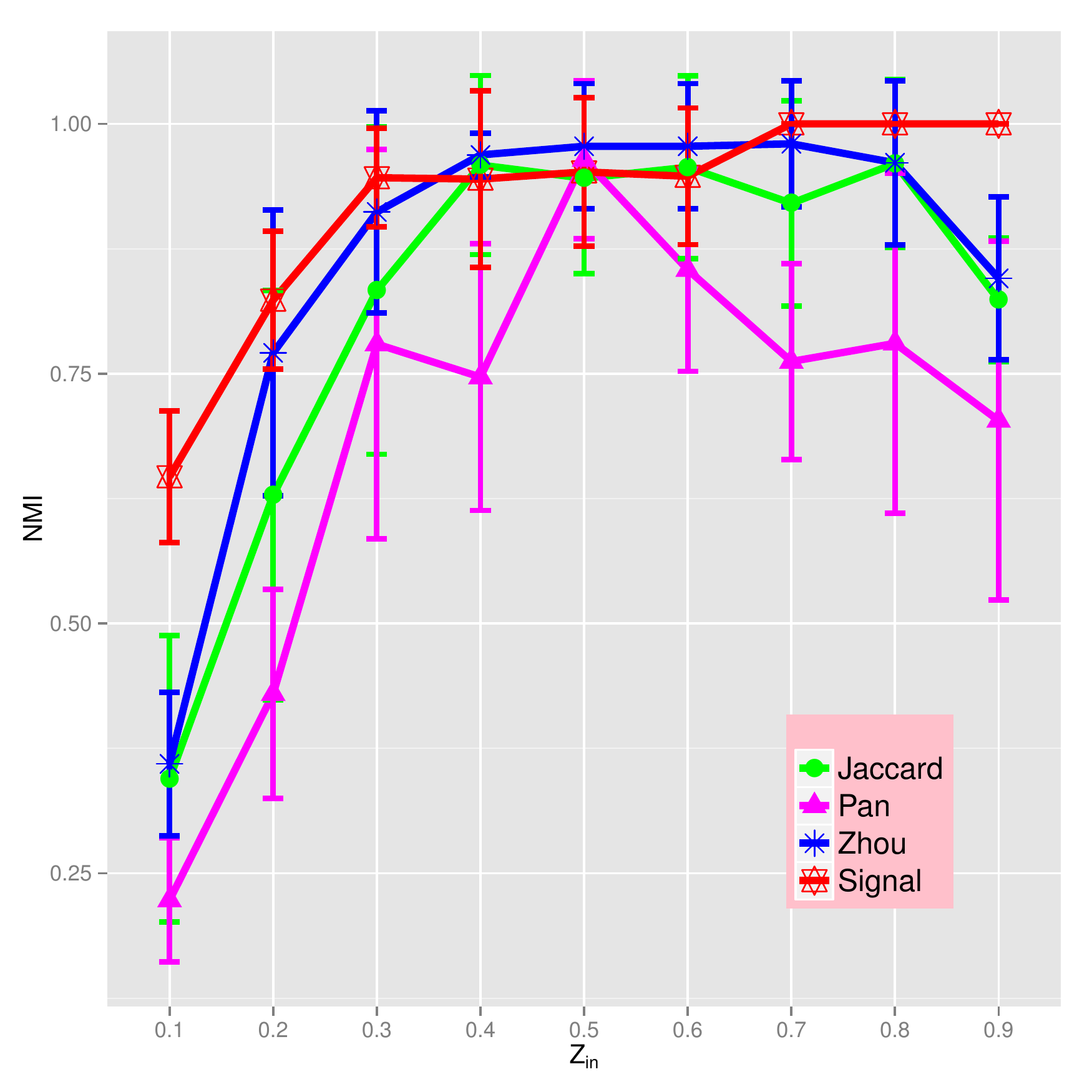}
		\hfill
		\includegraphics[width=0.45\linewidth]{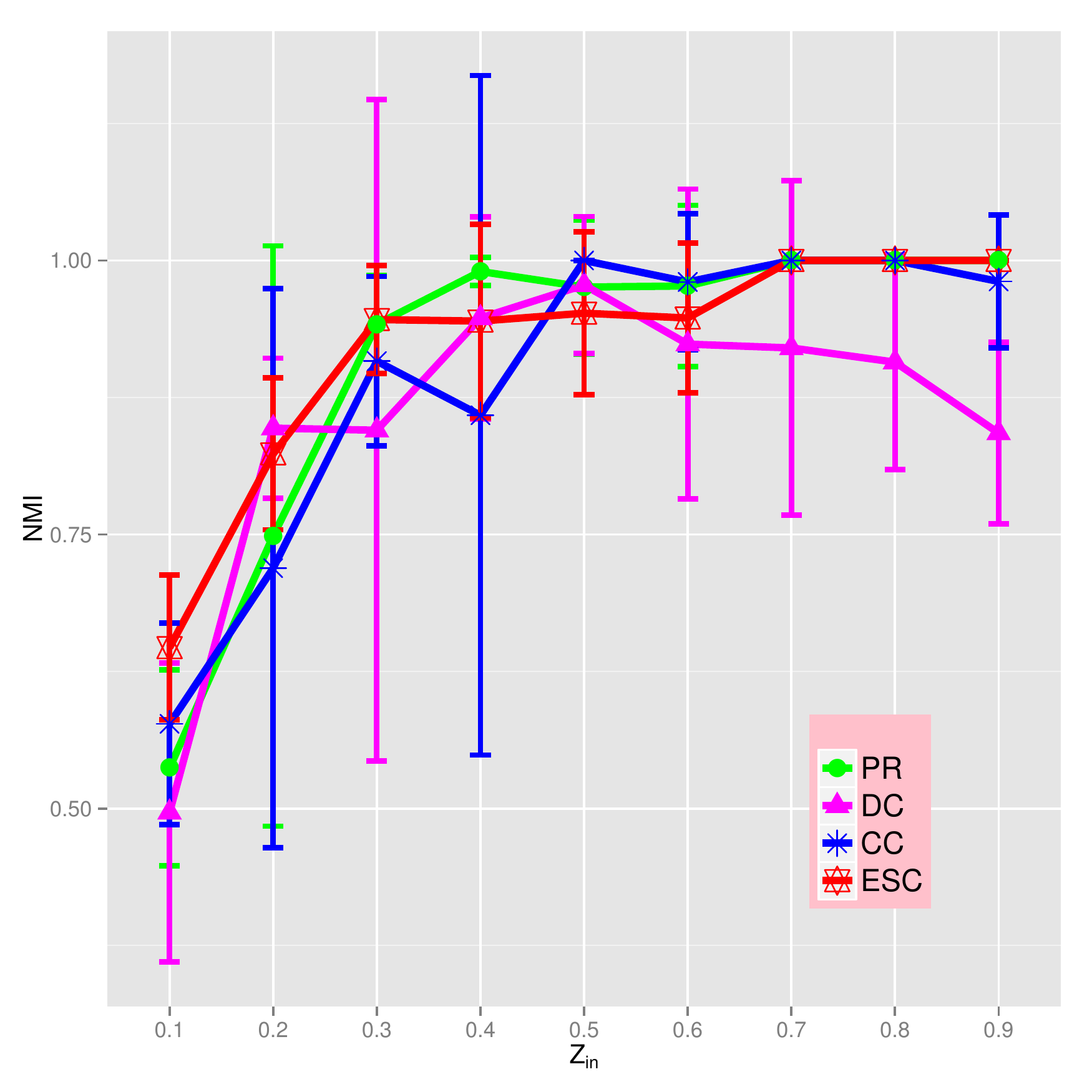}
		\hfill \parbox{.45\linewidth}{\centering\small a. Different similarity measures}
		\hfill \parbox{.45\linewidth}{\centering\small b. Different centrality measures}
 \hfill \caption{Comparison of similarity and centrality measures in the application of SMP  algorithm. Average NMI values (plus and minus one standard deviation) for 20 repeated experiments, as a function
of the average degree.} \label{sim_cen_dif_method} \end{figure} \end{center}

It is noteworthy that in the application of SMP algorithm, different similarity  and centrality measures could be adopted instead of the signal similarity and evidential semi-local centrality suggested in this paper. When using ESC for calculating the centrality, results by four different similarity metrics, {\em i.e.,} signal similarity, the simple Jaccard index and the measures proposed by \citet{pan2010detecting} (denoted by Pan in the figure) and \citet{zhou2009predicting}  (denoted by Zhou in the figure), are shown in Figure~\ref{sim_cen_dif_method}-a. As can be seen from the figure, the results by signal similarity  are better than the other indices in terms of NMI values. Here we could conclude that global similarity measures like  signal similarity are more applicable  for SMP than local ones. Figure~\ref{sim_cen_dif_method}-b depicts the behavior of SMP with difference centrality measures but the same (signal) similarity index. It can be seen that ESC and PR are better among the four measures, {\em i.e.,} ESC, PageRank (PR) \citep{brin1998anatomy}, Degree Centrality (DC), and Closeness Centrality (CC). Although there is no significant difference between ESC and PR,  the performance of ESC is more stable than PR. This paper is not focusing on the comparison of different similarity and centrality measures, thus in the following experiment we only consider the signal similarity and evidential semi-local centrality.

\begin{center} \begin{figure}[!thbt] \centering
		\includegraphics[width=0.45\linewidth]{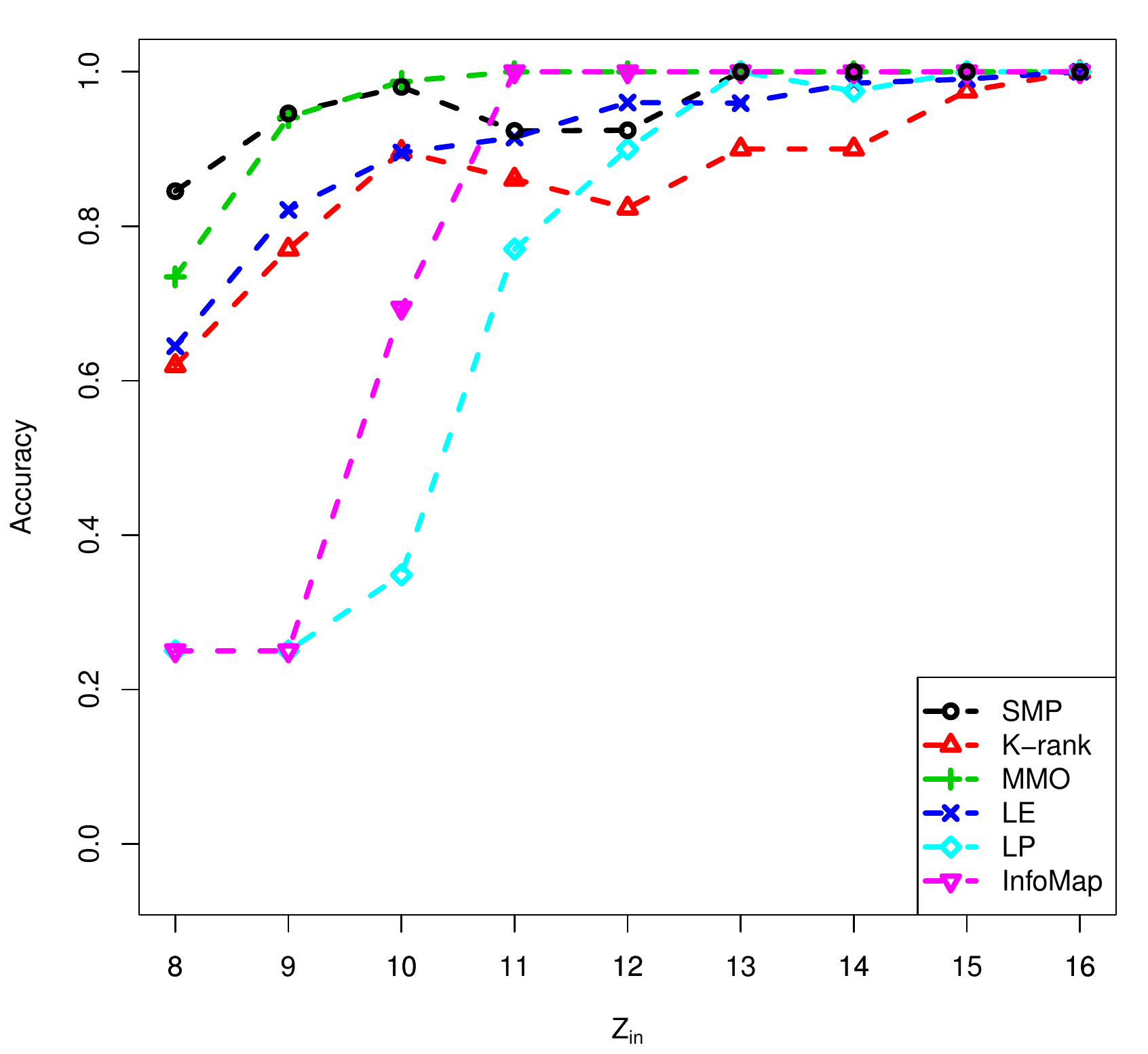}
		\hfill
		\includegraphics[width=0.45\linewidth]{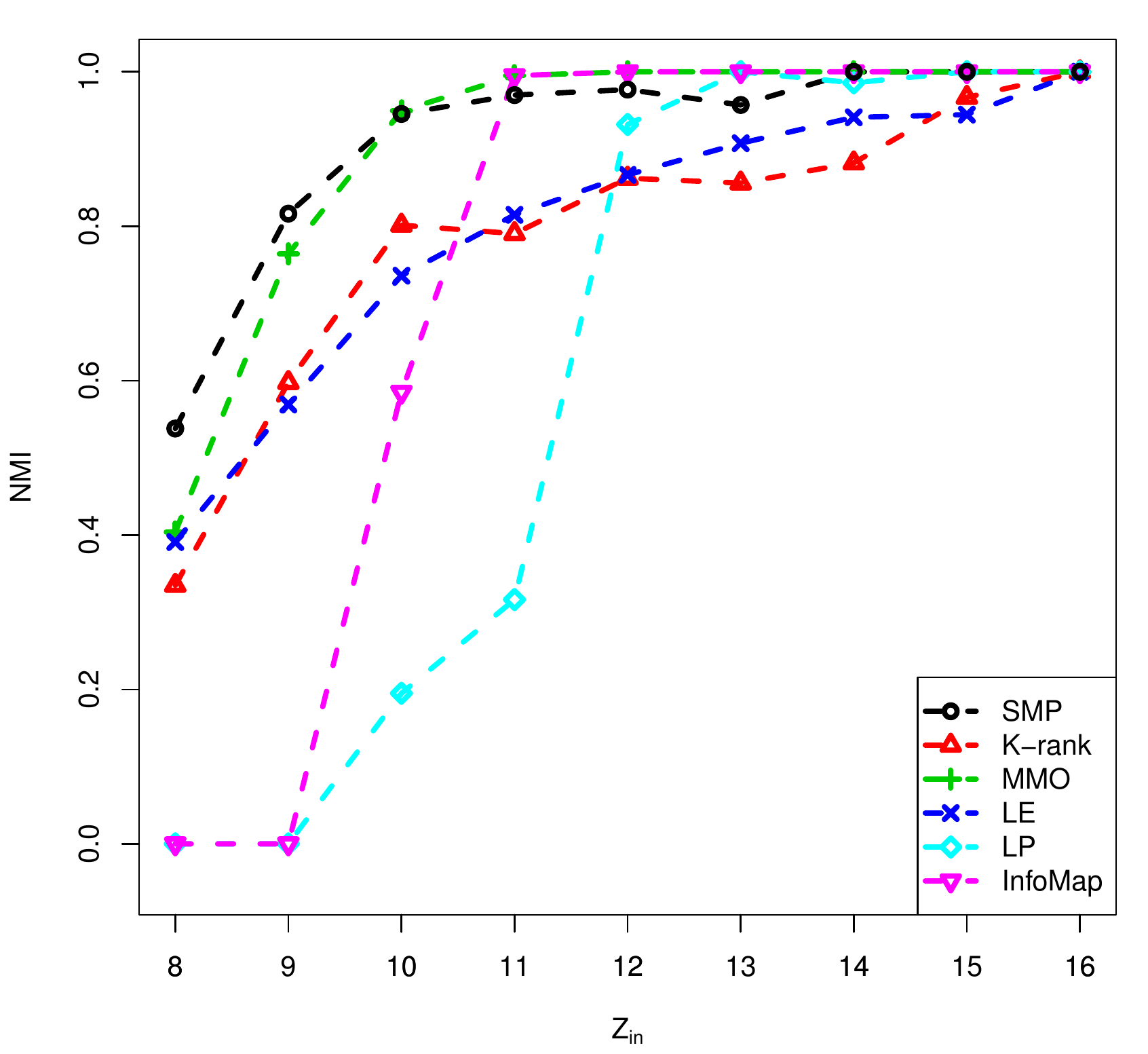}
		\hfill \parbox{.45\linewidth}{\centering\small a. Accuracy}
		\hfill \parbox{.45\linewidth}{\centering\small b. NMI}
 \hfill \caption{Comparison of SMP and other algorithms in Girvan and Newman's networks.} \label{simulate_network_com_dif_method} \end{figure} \end{center}

For each  $Z_\text{in}$, the experiment is repeated 20 times and the mean values of the evaluating measures are
reported. The average values of  the indices by accuracy and NMI  using SMP and the other five  algorithms  with different values of $Z_\text{in}$  are displayed in Figure~\ref{simulate_network_com_dif_method}-a and Figure~\ref{simulate_network_com_dif_method}-b respectively. The results show that in terms of  accuracy and NMI, all the methods perform well when $Z_\text{in}$ is large. However, when $Z_\text{in}$ is smaller than 10, they have different performances.  LP and InfoMap have the worst results as they could not work when $Z_\text{in} <10$. SMP and MMO are best in general among all the methods. Although MMO is superior to SMP when $Z_\text{in}= 11$ and $Z_\text{in}=12$, the superiority is not obvious. SMP is significantly better than MMO when $Z_\text{in}$ is small (especially when $Z_\text{in} = 8$). Moreover,   with the decreasing of $Z_\text{in}$, the performance of
SMP  does not drop so dramatically as the case in other methods. This demonstrates that using multiple members with various prototype weights is able to characterize the structure of clusters more precisely no matter whether the network has clear community structure or not, which in turn helps to produce a partition of the graph with good quality.

\begin{center} \begin{figure}[!thbt] \centering
		\includegraphics[width=0.45\linewidth]{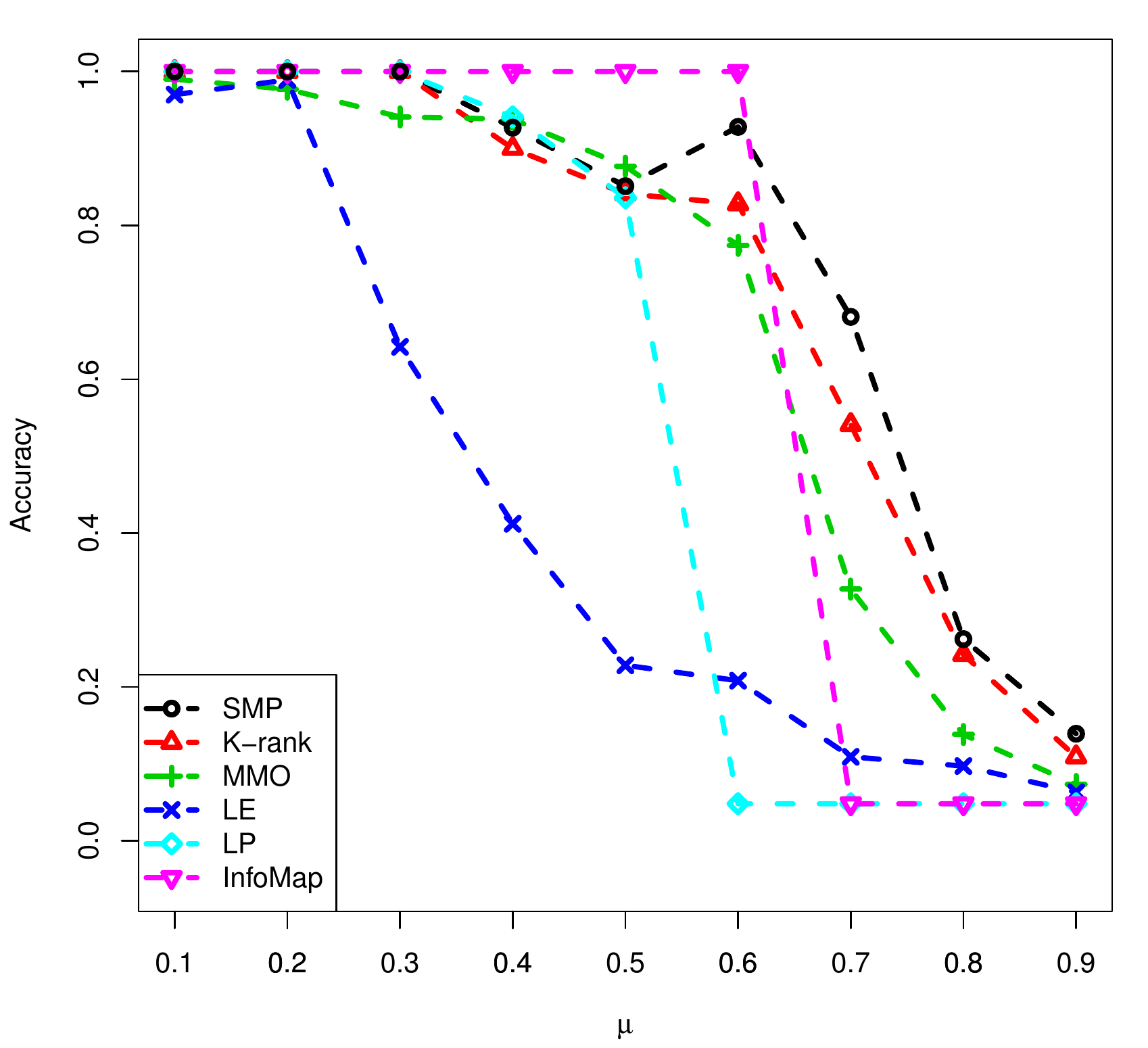}
		\hfill
		\includegraphics[width=0.45\linewidth]{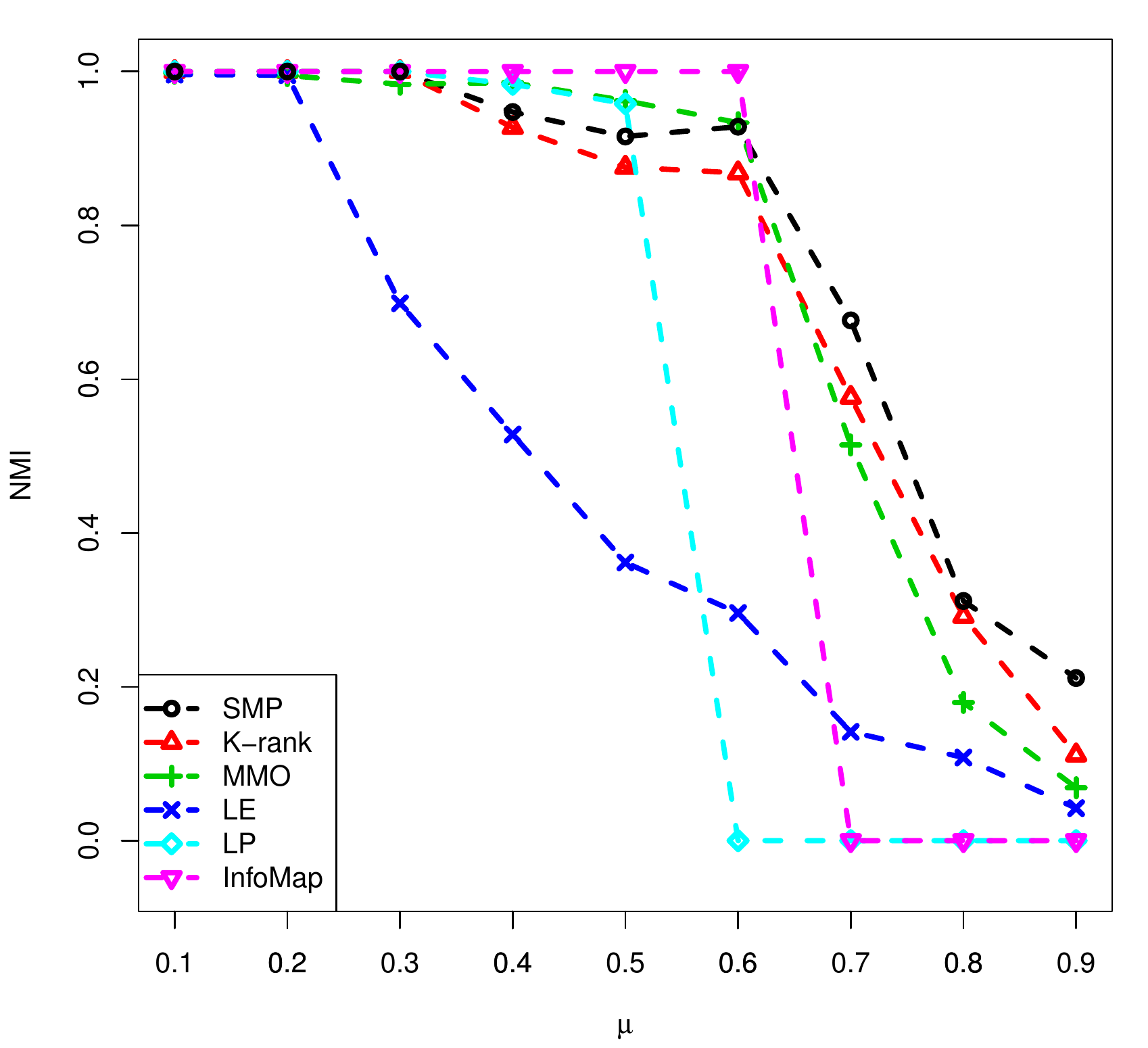}
		\hfill \parbox{.45\linewidth}{\centering\small a. Accuracy}
		\hfill \parbox{.45\linewidth}{\centering\small b. NMI}
 \hfill \caption{Comparison of SMP and other algorithms in LFR networks.  The number of nodes is $N=1000$.  The average degree is $|k| = 20$, and the pair for the exponents is $(\gamma, \beta)=(2,1)$.} \label{LFR1000} \end{figure} \end{center}

\begin{center} \begin{figure}[!thbt] \centering
		\includegraphics[width=0.45\linewidth]{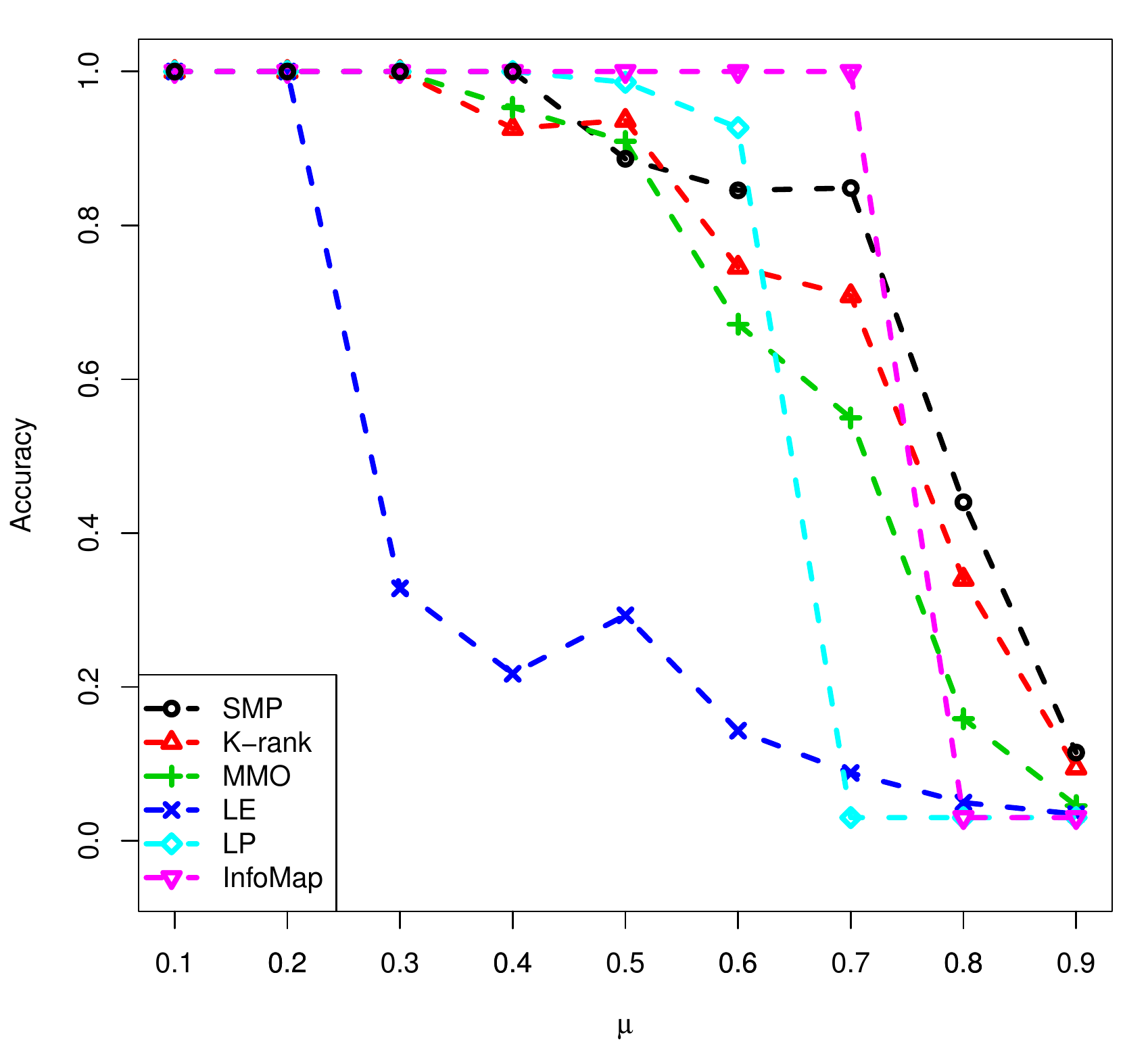}
		\hfill
		\includegraphics[width=0.45\linewidth]{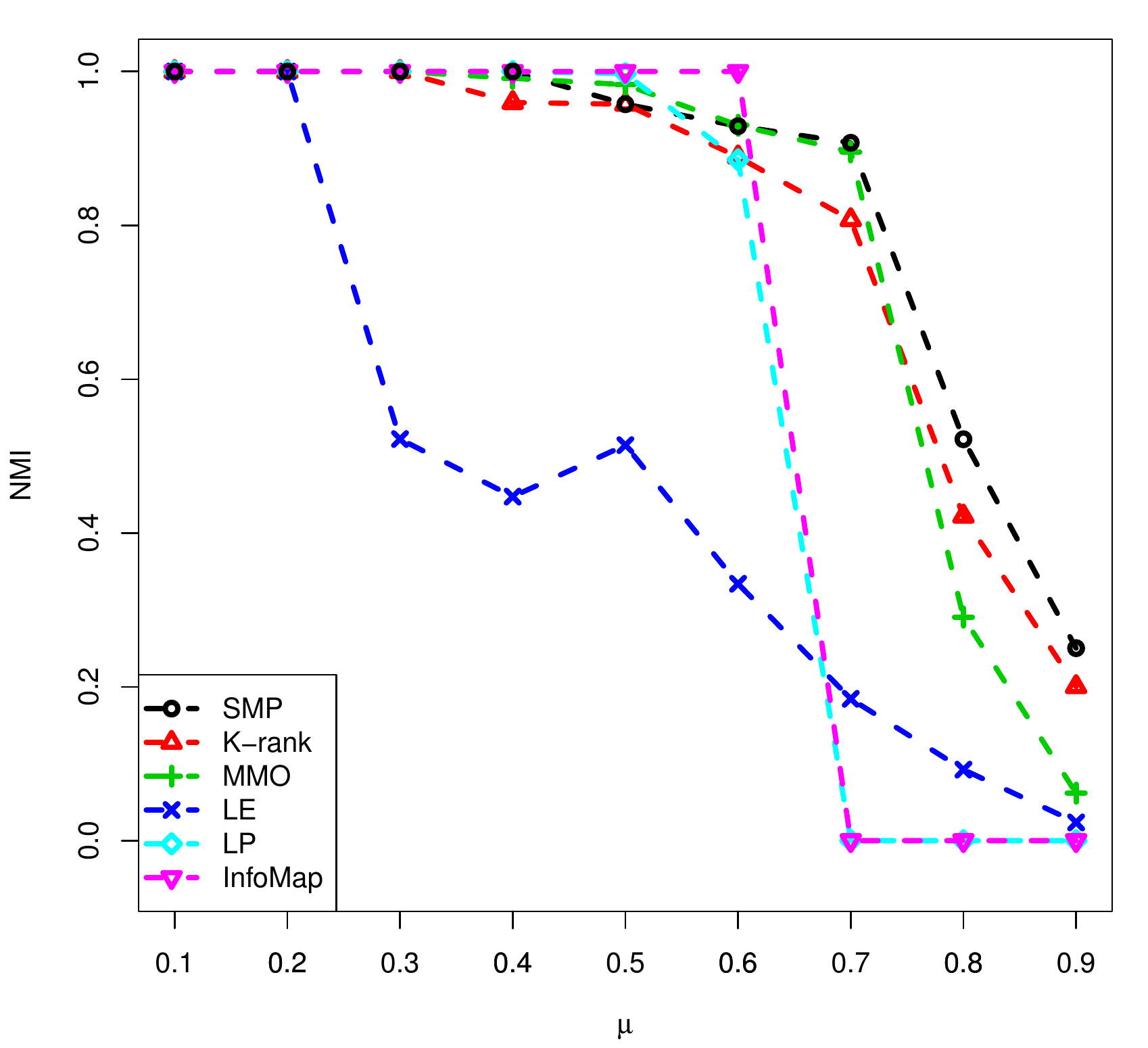}
		\hfill \parbox{.45\linewidth}{\centering\small a. Accuracy}
		\hfill \parbox{.45\linewidth}{\centering\small b. NMI}
 \hfill \caption{Comparison of SMP and other algorithms in LFR networks.  The number of nodes is $N=2000$. The average degree is $|k| = 30$, and the pair for the exponents is $(\gamma, \beta)=(2,1)$.} \label{LFR2000} \end{figure} \end{center}

\begin{center} \begin{figure}[!thbt] \centering
		\includegraphics[width=0.45\linewidth]{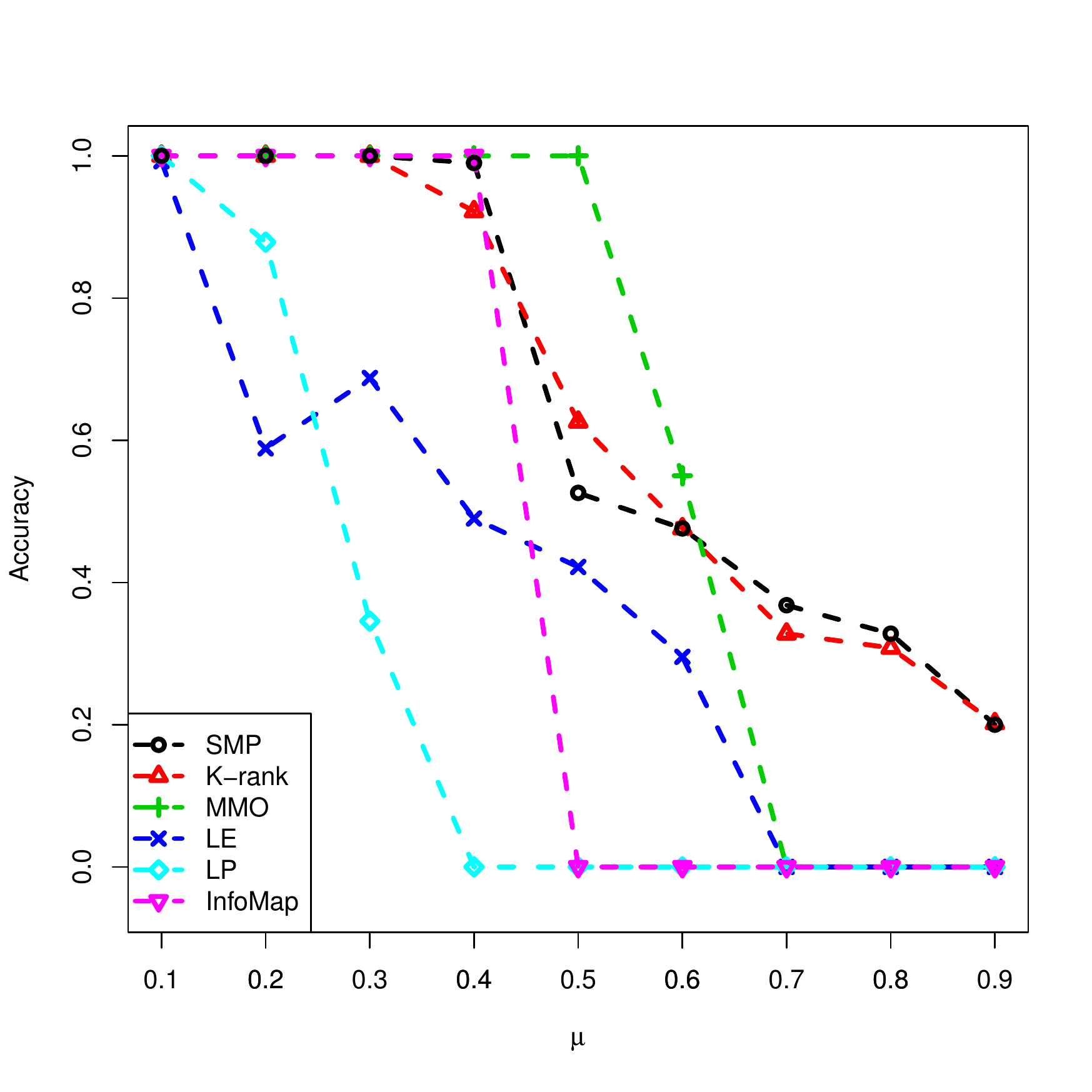}
		\hfill
		\includegraphics[width=0.45\linewidth]{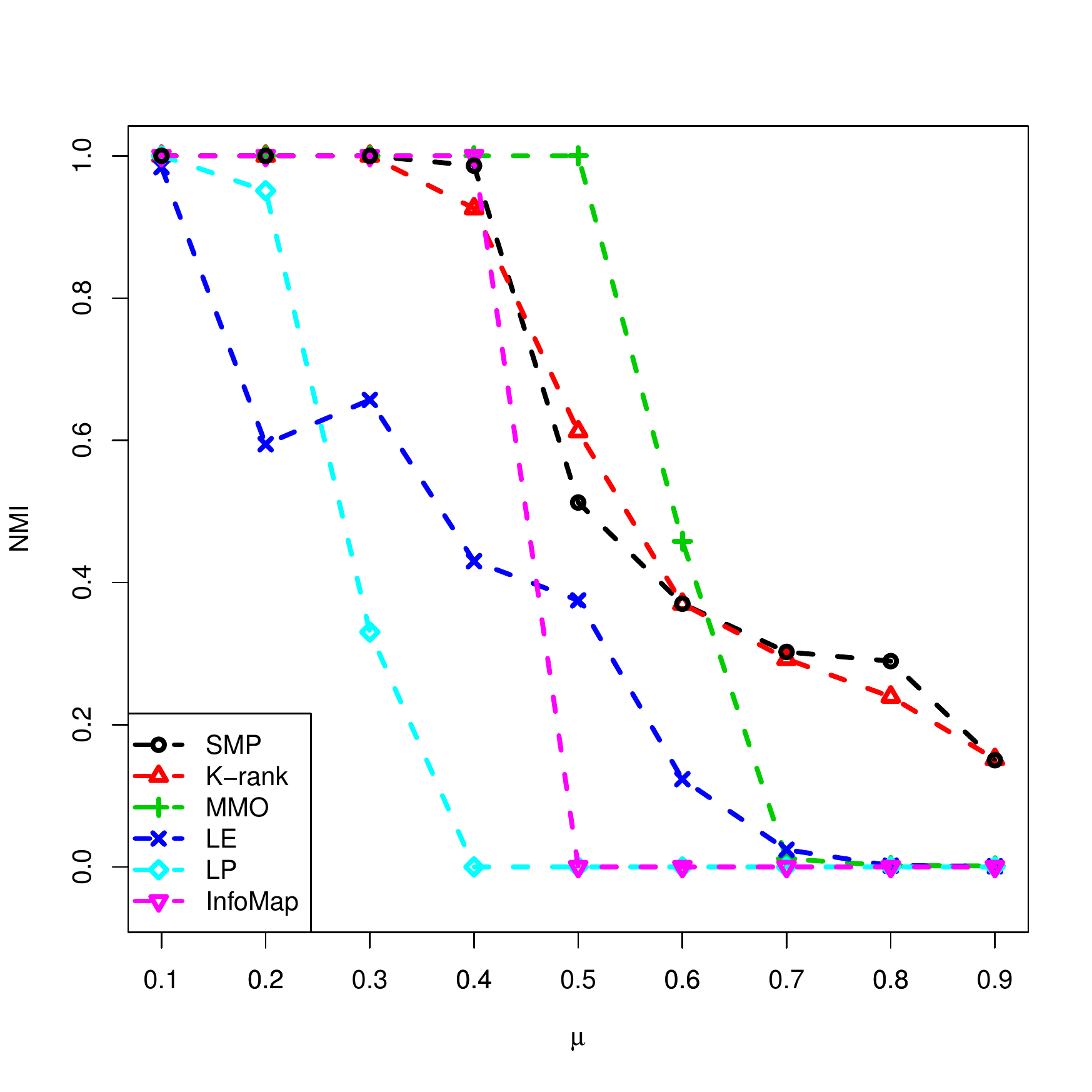}
		\hfill \parbox{.45\linewidth}{\centering\small a. Accuracy}
		\hfill \parbox{.45\linewidth}{\centering\small b. NMI}
 \hfill \caption{Comparison of SMP and other algorithms in LFR networks.  The number of nodes is $N=5000$. The average degree is $|k| = 30$, and the pair for the exponents is $(\gamma, \beta)=(2,1)$.} \label{LFR5000} \end{figure} \end{center}

The LFR benchmark network \citep{lancichinetti2008benchmark} is an artificial network for community detection, which is claimed to process some basic statistical properties found in real networks, such as heterogeneous distributions of degree and community size. The results of different methods in three kinds of LFR networks with 1000, 2000 and 5000 nodes are displayed in Figures~\ref{LFR1000}--\ref{LFR5000} respectively.  The parameter $\mu$ illustrated in  $x$-axis in the figures identifies  whether the network has clear communities. When $\mu$ is small, the graph has well community structure. In such a case, almost all the methods perform well. But we can see that when $\mu$ is large, the results by SMP have relatively  large values of NMI,  and the performance of SMP and $K$-rank do not drop dramatically as the case in other methods. SMP slightly outperforms $K$-rank especially when $\mu$ is large, this could be attributed to the multi-prototype representation of communities. Overall, from the two types of benchmarks, SMP fits for the networks no matter whether they have clear community structures or not.
\subsection{Real world networks}
\label{realdata}
\begin{center} \begin{figure}[!thbt] \centering
		\includegraphics[width=0.45\linewidth]{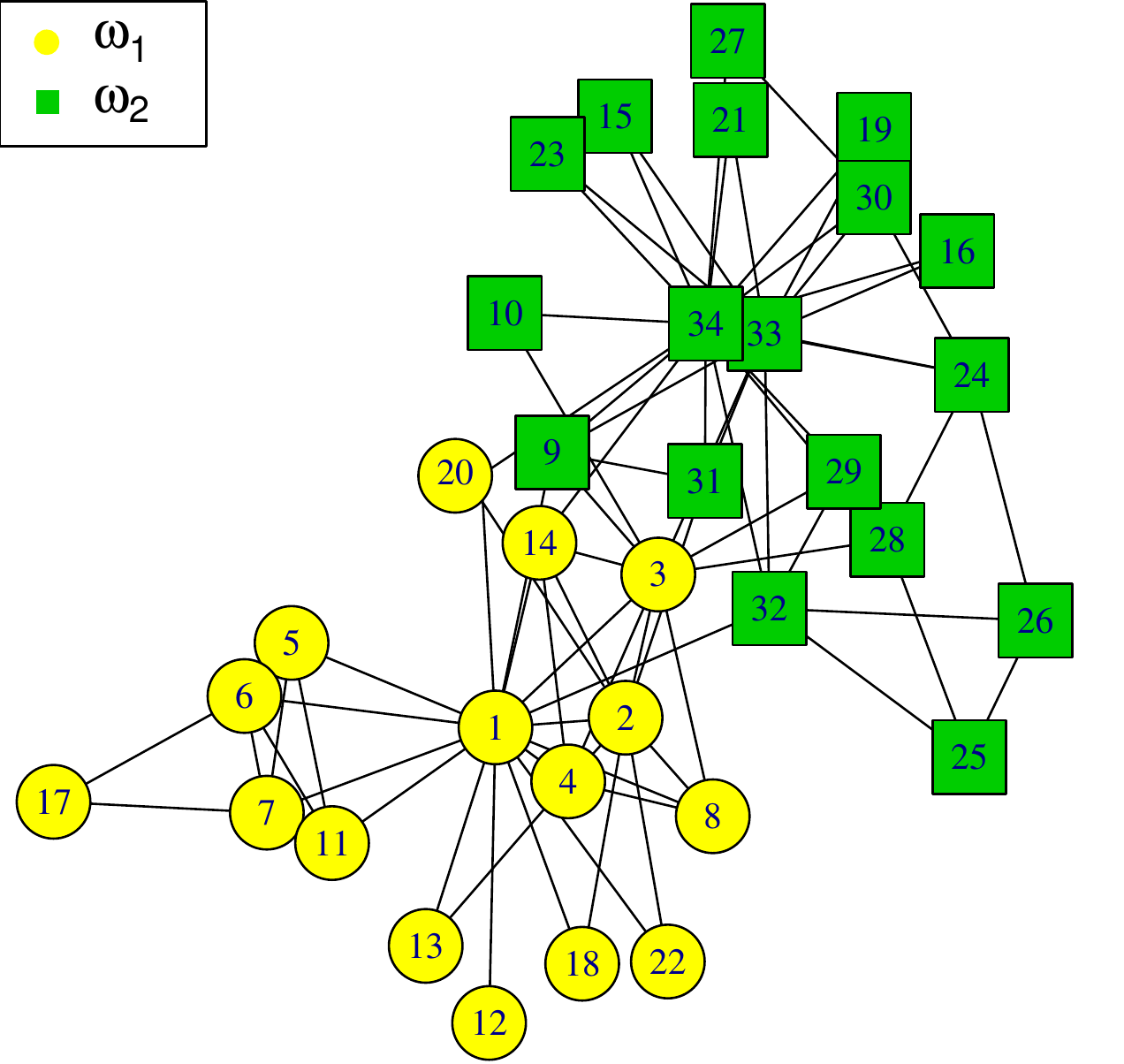}
		\hfill
		\includegraphics[width=0.45\linewidth]{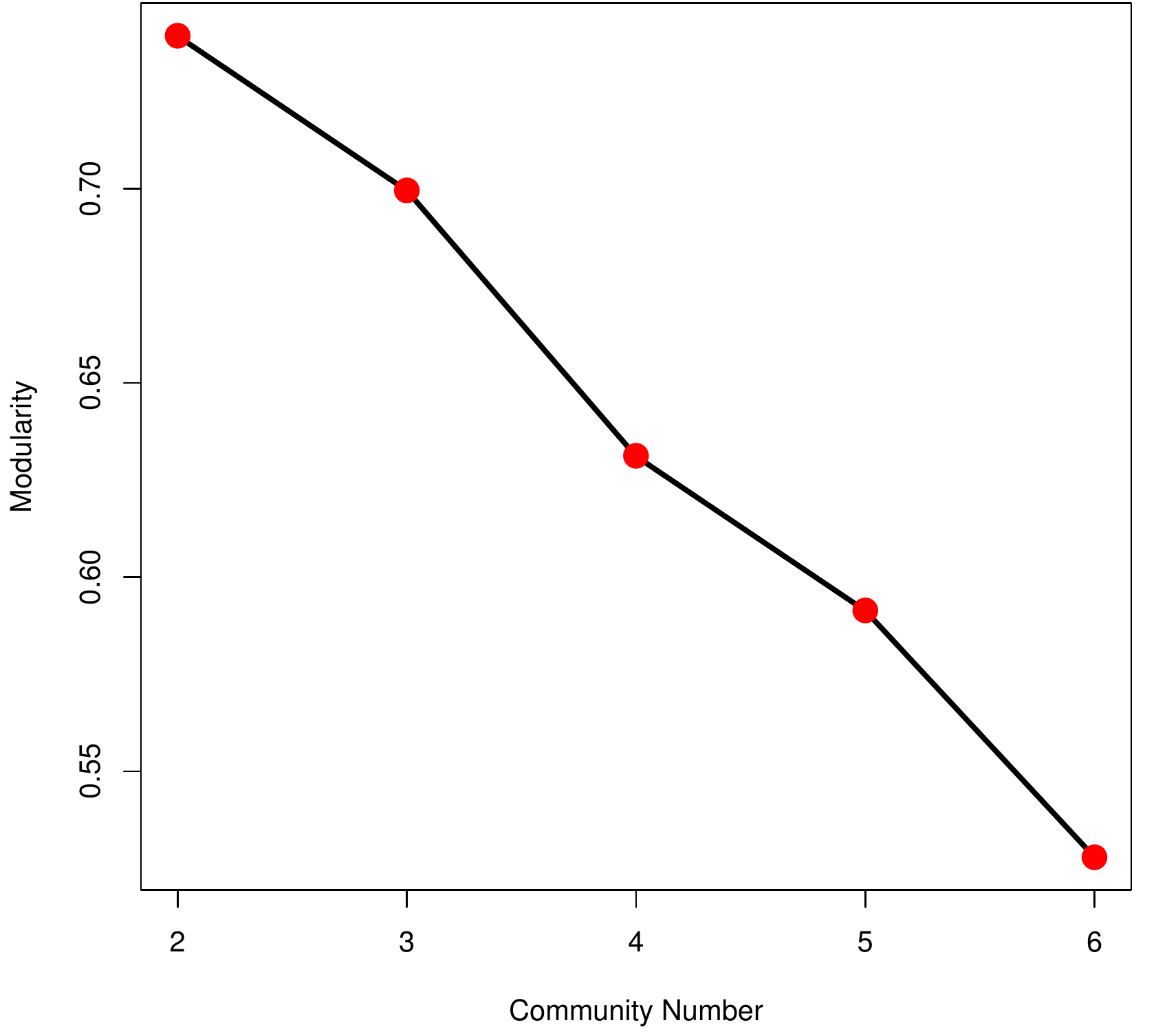}
		\hfill \parbox{.45\linewidth}{\centering\small a. Original Karate Club network}
		\hfill \parbox{.45\linewidth}{\centering\small b. Modularity function}
 \hfill \caption{The Karate Club network and the modularity values varying with community numbers.} \label{karate} \end{figure} \end{center}

 \textbf{A. Zachary's Karate Club.}
To evaluate the effectiveness of the proposed method applied on real-world networks, we first test on a widely used benchmark  in detecting community structures, ``Karate Club" \citep{zachary1977information},  studied by Wayne Zachary. The network consists of 34  nodes and 78 edges representing the friendship among the members of the club. During the development, a dispute arose between the club's administrator and instructor, which eventually resulted in the club  split into two smaller clubs, centered around the administrator and the instructor respectively (see Figure~\ref{karate}-a).

The values of the modularity with different number of communities are displayed in Figure~\ref{karate}-b. The modularity function peaks when $K=2$. This is in consistent with the fact that the network has two groups. The discovered communities are illustrated in Table~\ref{karatetable}. The table also shows the prototype weights in each of the found group. As we can see, node 1  makes the most contribution to community 1, while node 34 is most important to community 2. This confirms the center role of the two persons in their own communities. On the contrary, nodes 17 and 25 seem not very important in their group in terms of their prototype weights. We can see that in Figure~\ref{karate}-a, these two nodes locate in the marginal parts. Therefore, the proposed SMP detection approach enables us to have a better understanding of the graph structure with the help of prototype weights.
\begin{table}[ht]
 \centering \caption{The results for Karate Club network. The notation $u_{ij}$ denotes the fuzzy membership of node $n_i$ to community $j$, and PW is short for prototype weights. The nodes are order by prototype weights in each community.}
\begin{tabular}{rrrr|rrrr}
\hline
\multicolumn{4}{c|}{Community 1}& \multicolumn{4}{c}{Community 2}\\
  \hline
Node ID & $u_{i1}$ & $u_{i2}$ &PW & Node ID &$u_{i1}$& $u_{i2}$& PW \\
  \hline
1 & 0.5324 & 0.4676 & 0.1166 & 34 & 0.4607 & 0.5393 & 0.1025 \\
  2 & 0.5305 & 0.4695 & 0.0929 & 33 & 0.4582 & 0.5418 & 0.0940 \\
  4 & 0.5385 & 0.4615 & 0.0881 & 24 & 0.4469 & 0.5531 & 0.0738 \\
  3 & 0.5091 & 0.4909 & 0.0857 & 32 & 0.4798 & 0.5202 & 0.0698 \\
  8 & 0.5404 & 0.4596 & 0.0786 & 30 & 0.4424 & 0.5576 & 0.0679 \\
  14 & 0.5175 & 0.4825 & 0.0786 & 9 & 0.4882 & 0.5118 & 0.0595 \\
  6 & 0.5576 & 0.4424 & 0.0536 & 31 & 0.4772 & 0.5228 & 0.0595 \\
  7 & 0.5576 & 0.4424 & 0.0536 & 15 & 0.4464 & 0.5536 & 0.0532 \\
  18 & 0.5486 & 0.4514 & 0.0524 & 16 & 0.4464 & 0.5536 & 0.0532 \\
  20 & 0.5109 & 0.4891 & 0.0524 & 19 & 0.4464 & 0.5536 & 0.0532 \\
  22 & 0.5486 & 0.4514 & 0.0524 & 21 & 0.4464 & 0.5536 & 0.0532 \\
  5 & 0.5564 & 0.4436 & 0.0488 & 23 & 0.4464 & 0.5536 & 0.0532 \\
  11 & 0.5564 & 0.4436 & 0.0488 & 28 & 0.4707 & 0.5293 & 0.0474 \\
  13 & 0.5513 & 0.4487 & 0.0476 & 29 & 0.4788 & 0.5212 & 0.0408 \\
  12 & 0.5488 & 0.4512 & 0.0334 & 27 & 0.4420 & 0.5580 & 0.0392 \\
  17 & 0.5734 & 0.4266 & 0.0164 & 10 & 0.4802 & 0.5198 & 0.0307 \\
   &  &  &  & 26 & 0.4582 & 0.5418 & 0.0268 \\
   &  & &  & 25 & 0.4671 & 0.5329 & 0.0223 \\
   \hline
\end{tabular}\label{karatetable}
\end{table}

\begin{center} \begin{figure}[!thbt] \centering
		\includegraphics[width=0.45\linewidth]{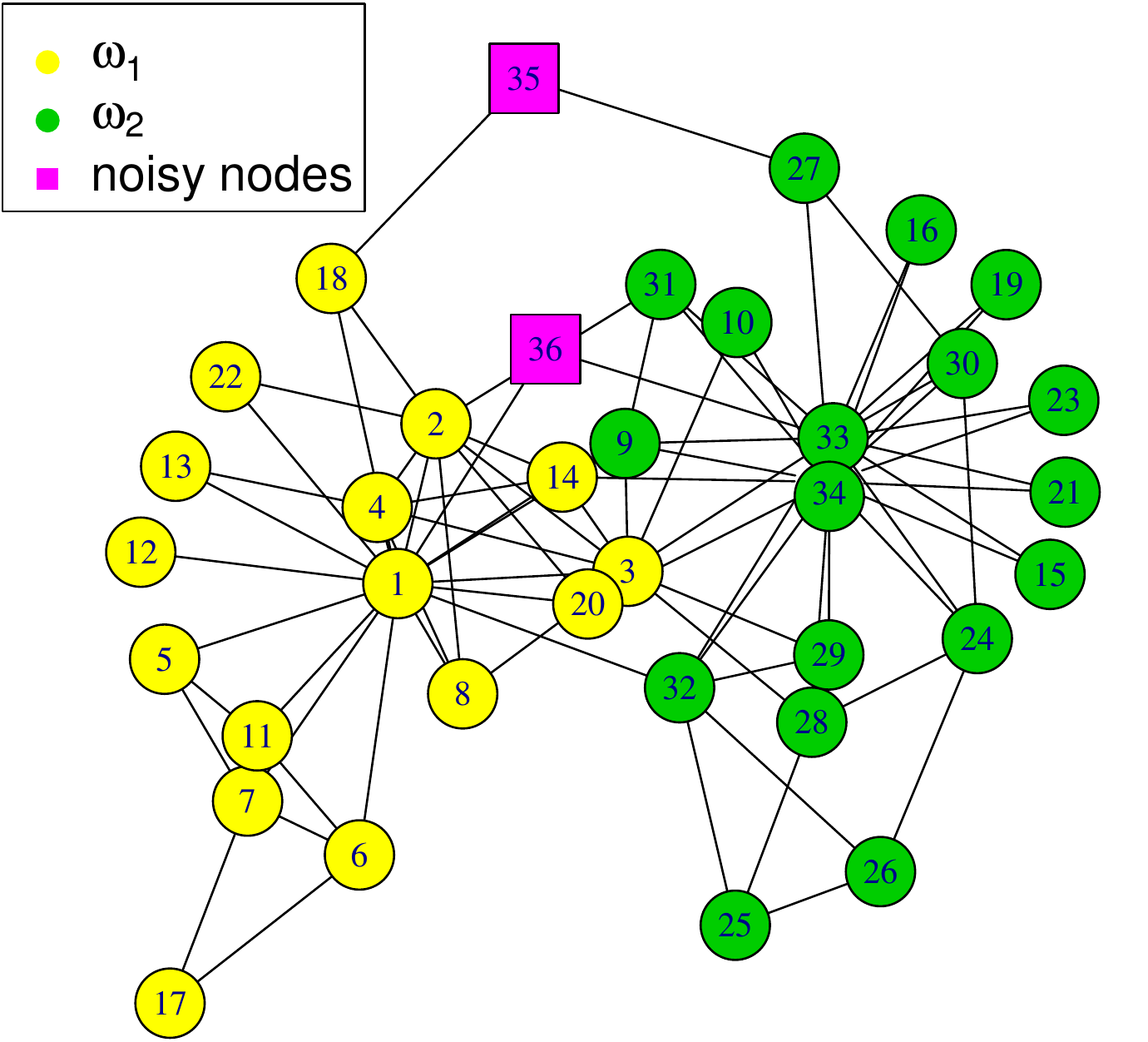}
		\hfill
		\includegraphics[width=0.45\linewidth]{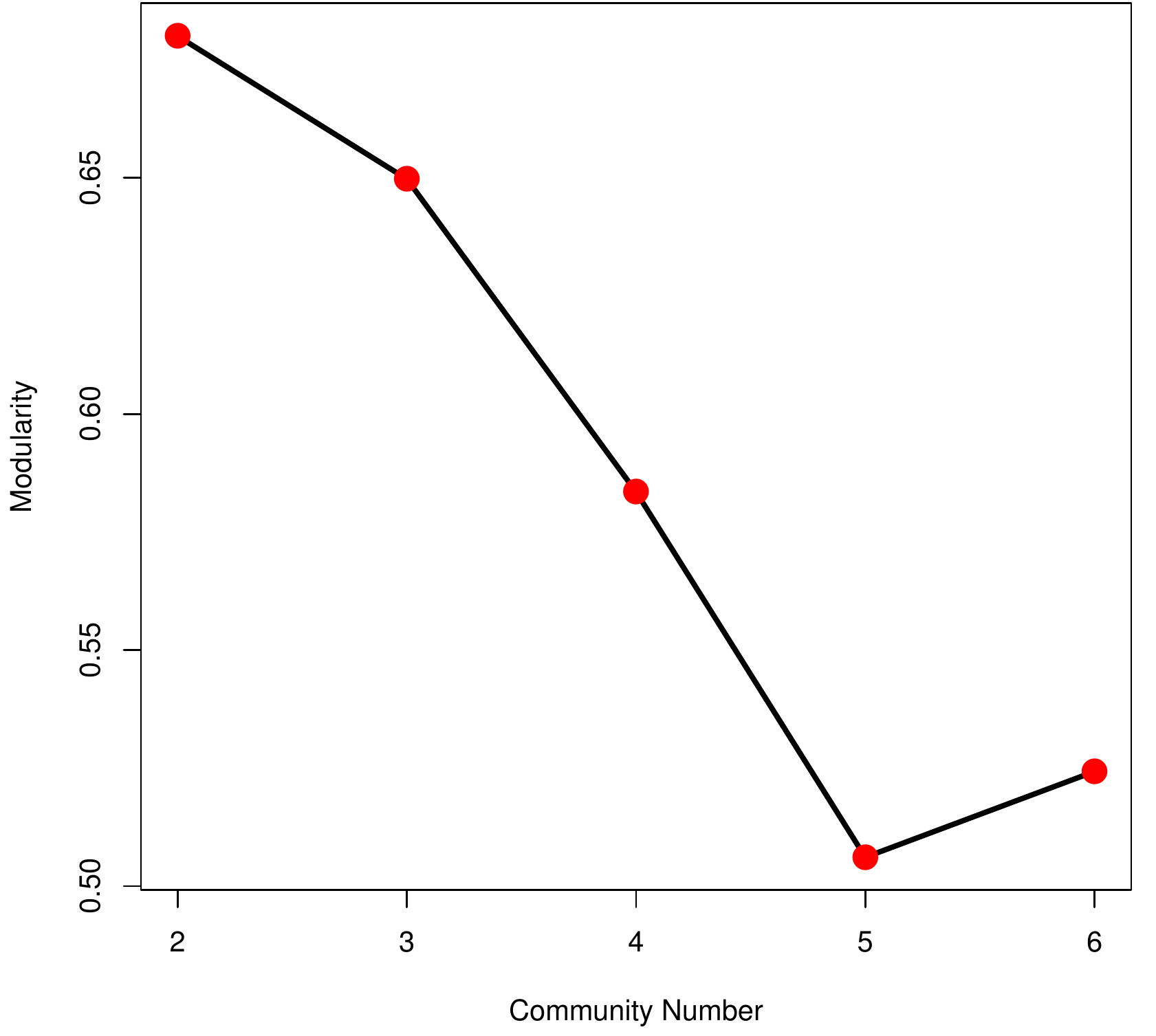}
		\hfill \parbox{.45\linewidth}{\centering\small a. Karate Club network with two added nodes}
		\hfill \parbox{.45\linewidth}{\centering\small b. Modularity function}
 \hfill \caption{The Karate Club network with added nodes and the modularity values varying with community numbers.} \label{karateadd} \end{figure} \end{center}

\textbf{B. Karate Club network with some added noisy nodes.} In this test, two noisy nodes are added to the original
Karate Club network (see Figure \ref{karateadd}-a). The first one is node 35,
which is directly connected with nodes 18 and 27. The other one is 36, which
is connected to nodes 1 and 33. It can be seen that node 36 has stronger relationships  with both communities than
node 35. This is due to the fact that the nodes connected to node 36
play leader roles in their own groups, but node 35 contacts with two
marginal nodes which have only ``small" or insignificant  roles in their own groups. The modularity values varying with different community numbers are depicted in Figure \ref{karateadd}-b and the detected results are displayed in Table \ref{karateaddtable}.

From Table \ref{karateaddtable} we can see that the fuzzy membership values of nodes 35 and 36 are almost the same  for both communities (approximatively equal to 0.5). These results could not reflect the difference between ignorance and uncertainty. As
node 35 is only related to one outward node of each  community, thus we are
ignorant about which community it really belongs to, or we say node 35 is an outlier. On the contrary,
node 36 connects with the key members (playing an important role in the
community) in both communities.
Thus there is uncertainty rather than ignorance about which community node 36
is in. In this network, node 36 is a ``good" member for both communities,
whereas node 35 is a ``poor" member.  As mentioned before, the inability to distinguish the outliers from the uncertain nodes with equal memberships is caused by the relative similarity used in fuzzy memberships. In SMP, the prototype weights could be utilized to solve this problem and to detect the outliers. As shown in Table \ref{karateaddtable}, the prototype weight of node 35 is the least in the community, but node 36 contributes much more than node 35. Therefore, node 35 has no contribution to both communities (the prototype weight of node 35 for community 1 is 0.0052, and 0 for   community 2), and it could be recognized as an outlier. This example further demonstrates the fact that prototype weights indeed enable us to gain a better understanding of the graph structure, especially for detecting outliers in the network.
\begin{table}[ht]
\centering \caption{The results for Karate Club network with added nodes. The notation $u_{ij}$ denotes the fuzzy membership of node $n_i$ to community $j$, and PW is short for prototype weights. The nodes are order by PW in each community.}
\begin{tabular}{rrrr|rrrr}
\hline
\multicolumn{4}{c|}{Community 1}& \multicolumn{4}{c}{Community 2}\\
  \hline
Node ID & $u_{i1}$ & $u_{i2}$ &PW & Node ID &$u_{i1}$& $u_{i2}$& PW \\
  \hline
1 & 0.5278 & 0.4722 & 0.1111 & 34 & 0.4656 & 0.5344 & 0.1028 \\
  2 & 0.5271 & 0.4729 & 0.0888 & 33 & 0.4651 & 0.5349 & 0.0944 \\
  4 & 0.5344 & 0.4656 & 0.0836 & 24 & 0.4534 & 0.5466 & 0.0737 \\
  3 & 0.5084 & 0.4916 & 0.0814 & 32 & 0.4824 & 0.5176 & 0.0696 \\
  8 & 0.5360 & 0.4640 & 0.0747 & 30 & 0.4506 & 0.5494 & 0.0680 \\
  14 & 0.5158 & 0.4842 & 0.0747 & 9 & 0.4899 & 0.5101 & 0.0598 \\
  18 & 0.5399 & 0.4601 & 0.0528 & 31 & 0.4801 & 0.5199 & 0.0598 \\
  6 & 0.5511 & 0.4489 & 0.0520 & 15 & 0.4533 & 0.5467 & 0.0534 \\
  7 & 0.5511 & 0.4489 & 0.0520 & 16 & 0.4533 & 0.5467 & 0.0534 \\
  20 & 0.5099 & 0.4901 & 0.0506 & 19 & 0.4533 & 0.5467 & 0.0534 \\
  22 & 0.5427 & 0.4573 & 0.0506 & 21 & 0.4533 & 0.5467 & 0.0534 \\
  5 & 0.5498 & 0.4502 & 0.0475 & 23 & 0.4533 & 0.5467 & 0.0534 \\
  11 & 0.5498 & 0.4502 & 0.0475 & 28 & 0.4740 & 0.5260 & 0.0471 \\
  13 & 0.5454 & 0.4546 & 0.0462 & 29 & 0.4813 & 0.5187 & 0.0404 \\
  12 & 0.5427 & 0.4573 & 0.0330 & 27 & 0.4539 & 0.5461 & 0.0395 \\
  36 & 0.5016 & 0.4984 & 0.0330 & 10 & 0.4826 & 0.5174 & 0.0309 \\
  17 & 0.5658 & 0.4342 & 0.0154 & 26 & 0.4628 & 0.5372 & 0.0258 \\
  35 & 0.5020 & 0.4980 & 0.0052 & 25 & 0.4705 & 0.5295 & 0.0212 \\
   \hline
\end{tabular} \label{karateaddtable}
\end{table}

We also test our method on four other real-world graphs: American football network, Dolphins network, Lesmis network and Political
books network\footnote{These data sets can be found in http://networkdata.ics.uci.edu/index.php}. 
The values of the two indices, accuracy and NMI, applied to evaluate the performance of different methods are listed in Table~\ref{real} and Table~\ref{real1} respectively\footnote{All these real-world graphs are with known community structure, thus the accuracy and NMI are calculated based on the ground truth and the partition got by different algorithms.}. It can been seen from the tables,  SMP application results in a community structure with highest accuracy level in most cases.  In terms of the performance measure NMI, SMP also outperforms the other algorithms. It should be noted that some methods provide partitions with high accuracy but low NMI. This may be caused by the fact that they cluster the nodes  into too many small communities. The partition rules of both $K$-rank and SMP are based on node similarity. These two approaches are better than the others in general,  and the effectiveness could be attributed to the high performance of vertex similarities. But the reason that SMP works better than $K$-rank in these real-world networks is largely because of the application of multiple prototype representation of communities.

From the above extensive experimental results, we can summarize the  compelling properties of SMP as follows:
\begin{enumerate}[1)]
     \item In the partition process, SMP uses multiple prototypes to represent the communities. This is a useful extension of the existing community detection methods where only one prototype is allowed, especially when the analyzed graph has some complex community structures.
     \item  The prototype weights, as a by-product of the detection results, provide us with some valuable information about the community structure from another point of view, and enable us to gain a better understanding of the analyzed graph.
     \item SMP works well even for the graphs without clear community structures. It could avoid the problem of inability to distinguish the outliers from uncertain data  for fuzzy membership.
     \item Last but not the least, the experiments on both synthetic
           and real-world graph data sets demonstrate that  the proposed approach is a competitive candidate for community detection tasks compared with other five existing methods.
\end{enumerate}

\begin{table}[ht]
\centering \caption{Comparison of SMP and other algorithms by accuracy in real-world networks.}
\begin{tabular}{rrrrrr}
  \hline
& Karate & Football & Dolphins & Lesmis & Books \\
  \hline
 SMP & \textbf{1.0000} & \textbf{0.9345} & \textbf{1.0000} & 0.7792 & \textbf{0.8667} \\
  $K$-rank & \textbf{1.0000} & 0.9320 & \textbf{1.0000} & 0.8052 & 0.8537 \\
  MMO & \textbf{1.0000} & 0.8000 & 0.9516 & 0.7922 & 0.7276 \\
  LE & \textbf{1.0000} & 0.6261 & 0.9677 & 0.7273 & 0.8476 \\
  LP & 0.9706 & 0.9043 & \textbf{1.0000} & 0.7273 & 0.8476 \\
  InfoMap & \textbf{1.0000} & 0.9043 & 0.9839 & \textbf{0.8701} & 0.7854 \\
   \hline
\end{tabular}\label{real}
\end{table}

\begin{table}[ht]
\centering \caption{Comparison of SMP and other algorithms by NMI in real-world networks.}
\begin{tabular}{rrrrrr}
  \hline
& Karate & Football & Dolphins & Lesmis & Books \\
  \hline
  SMP & \textbf{1.0000} & \textbf{0.9235} & \textbf{1.0000} & 0.7444 & \textbf{0.5938} \\
  $K$-rank & \textbf{1.0000} & 0.9211 & \textbf{1.0000} & 0.7818 & 0.5741 \\
  MMO & 0.6873 & 0.8550 & 0.4617 & 0.7551 & 0.5121 \\
  LE & 0.6552 & 0.6952 & 0.5094 & 0.7182 & 0.5201 \\
  LP & 0.8255 & 0.9095 & 0.8230 & 0.7381 & 0.5485 \\
  InfoMap & 0.8255 & 0.8937 & 0.5629 & \textbf{0.8198} & 0.4935 \\
   \hline
\end{tabular}\label{real1}
\end{table}

\section{Conclusion}
In this paper, a new type of similarity-based community detection algorithm called SMP is proposed.  SMP could find not only  communities of each node but also weighted representative members of each group.  In real world community detection problems, information on both community labels and internal structure of each of the detected communities are important. One distinctive characteristic of the proposed method is that each community is presented by multiple prototypes, rather than by  single one object. The experiments on synthetic networks  show the effectiveness of the proposed method and the tests on real-world networks have further pointed out our method preforms better  than the existing ones. The results show that the way of using prototype weights to represent a cluster enables SMP to capture the various types of community structures more precisely and completely hence improves the quality of the detected communities. Moreover, more detail information on the discovered clusters may be obtained with the help of  prototype weights.

In real applications, the signal similarity measure and ESC centrality  utilized in the work could be replaced by any other index. For instance, if we want to apply the method to directed networks, the similarity and centrality measures for directed networks could be adopted. Therefore, we intend to  study on the comparison of difference measures and on the application into directed networks in our future research work. Meanwhile, not only centrality but also more other factors should be considered for determining the  prototype weights. Hence the way to optimize the prototype weights using the available information as much as possible will also be included in our further study.
\section*{Acknowledgements}
The authors are grateful to the anonymous reviewers for all
their remarks which helped us to clarify and improve the quality
of this paper. This work was supported by the National
Natural Science Foundation of China (Nos.61135001, 61403310). The study of the
first author in France was supported by the China Scholarship Council.
\bibliographystyle{elsarticle-num-names}
\addcontentsline{toc}{section}{\refname}\bibliography{paperlist}

\begin{thebibliography}{39}
\providecommand{\natexlab}[1]{#1}
\providecommand{\url}[1]{\texttt{#1}}
\providecommand{\urlprefix}{URL }
\expandafter\ifx\csname urlstyle\endcsname\relax
  \providecommand{\doi}[1]{doi:\discretionary{}{}{}#1}\else
  \providecommand{\doi}[1]{doi:\discretionary{}{}{}\begingroup
  \urlstyle{rm}\url{#1}\endgroup}\fi
\providecommand{\bibinfo}[2]{#2}

\bibitem[{Fortunato(2010)}]{fortunato2010community}
\bibinfo{author}{S.~Fortunato}, \bibinfo{title}{Community detection in graphs},
  \bibinfo{journal}{Physics Reports}
  \bibinfo{volume}{486}~(\bibinfo{number}{3}) (\bibinfo{year}{2010})
  \bibinfo{pages}{75--174}.

\bibitem[{Costa et~al.(2011)Costa, Oliveira~Jr, Travieso, Rodrigues,
  Villas~Boas, Antiqueira, Viana, and Correa~Rocha}]{costa2011analyzing}
\bibinfo{author}{L.~d.~F. Costa}, \bibinfo{author}{O.~N. Oliveira~Jr},
  \bibinfo{author}{G.~Travieso}, \bibinfo{author}{F.~A. Rodrigues},
  \bibinfo{author}{P.~R. Villas~Boas}, \bibinfo{author}{L.~Antiqueira},
  \bibinfo{author}{M.~P. Viana}, \bibinfo{author}{L.~E. Correa~Rocha},
  \bibinfo{title}{Analyzing and modeling real-world phenomena with complex
  networks: a survey of applications}, \bibinfo{journal}{Advances in Physics}
  \bibinfo{volume}{60}~(\bibinfo{number}{3}) (\bibinfo{year}{2011})
  \bibinfo{pages}{329--412}.

\bibitem[{Girvan and Newman(2002)}]{girvan2002community}
\bibinfo{author}{M.~Girvan}, \bibinfo{author}{M.~E. Newman},
  \bibinfo{title}{Community structure in social and biological networks},
  \bibinfo{journal}{Proceedings of the National Academy of Sciences}
  \bibinfo{volume}{99}~(\bibinfo{number}{12}) (\bibinfo{year}{2002})
  \bibinfo{pages}{7821--7826}.

\bibitem[{Newman(2006{\natexlab{a}})}]{newman2006modularity}
\bibinfo{author}{M.~E. Newman}, \bibinfo{title}{Modularity and community
  structure in networks}, \bibinfo{journal}{Proceedings of the National Academy
  of Sciences} \bibinfo{volume}{103}~(\bibinfo{number}{23})
  (\bibinfo{year}{2006}{\natexlab{a}}) \bibinfo{pages}{8577--8582}.

\bibitem[{Zhang et~al.(2007)Zhang, Wang, and Zhang}]{zhang2007identification}
\bibinfo{author}{S.~Zhang}, \bibinfo{author}{R.-S. Wang},
  \bibinfo{author}{X.-S. Zhang}, \bibinfo{title}{Identification of overlapping
  community structure in complex networks using fuzzy c-means clustering},
  \bibinfo{journal}{Physica A: Statistical Mechanics and its Applications}
  \bibinfo{volume}{374}~(\bibinfo{number}{1}) (\bibinfo{year}{2007})
  \bibinfo{pages}{483--490}.

\bibitem[{Jiang et~al.(2013)Jiang, Jia, and Yu}]{jiang2012efficient}
\bibinfo{author}{Y.~Jiang}, \bibinfo{author}{C.~Jia}, \bibinfo{author}{J.~Yu},
  \bibinfo{title}{An efficient community detection method based on rank
  centrality}, \bibinfo{journal}{Physica A: Statistical Mechanics and its
  Applications} \bibinfo{volume}{392}~(\bibinfo{number}{9})
  (\bibinfo{year}{2013}) \bibinfo{pages}{2182--2194}.

\bibitem[{Newman and Girvan(2004)}]{newman2004finding}
\bibinfo{author}{M.~E. Newman}, \bibinfo{author}{M.~Girvan},
  \bibinfo{title}{Finding and evaluating community structure in networks},
  \bibinfo{journal}{Physical review E}
  \bibinfo{volume}{69}~(\bibinfo{number}{2}) (\bibinfo{year}{2004})
  \bibinfo{pages}{026113}.

\bibitem[{Blondel et~al.(2008)Blondel, Guillaume, Lambiotte, and
  Lefebvre}]{blondel2008fast}
\bibinfo{author}{V.~D. Blondel}, \bibinfo{author}{J.-L. Guillaume},
  \bibinfo{author}{R.~Lambiotte}, \bibinfo{author}{E.~Lefebvre},
  \bibinfo{title}{Fast unfolding of communities in large networks},
  \bibinfo{journal}{Journal of Statistical Mechanics: Theory and Experiment}
  \bibinfo{volume}{2008}~(\bibinfo{number}{10}) (\bibinfo{year}{2008})
  \bibinfo{pages}{P10008}.

\bibitem[{Clauset et~al.(2004)Clauset, Newman, and Moore}]{clauset2004finding}
\bibinfo{author}{A.~Clauset}, \bibinfo{author}{M.~E. Newman},
  \bibinfo{author}{C.~Moore}, \bibinfo{title}{Finding community structure in
  very large networks}, \bibinfo{journal}{Physical review E}
  \bibinfo{volume}{70}~(\bibinfo{number}{6}) (\bibinfo{year}{2004})
  \bibinfo{pages}{066111}.

\bibitem[{Duch and Arenas(2005)}]{duch2005community}
\bibinfo{author}{J.~Duch}, \bibinfo{author}{A.~Arenas},
  \bibinfo{title}{Community detection in complex networks using extremal
  optimization}, \bibinfo{journal}{Physical review E}
  \bibinfo{volume}{72}~(\bibinfo{number}{2}) (\bibinfo{year}{2005})
  \bibinfo{pages}{027104}.

\bibitem[{Fortunato and Barthelemy(2007)}]{fortunato2007resolution}
\bibinfo{author}{S.~Fortunato}, \bibinfo{author}{M.~Barthelemy},
  \bibinfo{title}{Resolution limit in community detection},
  \bibinfo{journal}{Proceedings of the National Academy of Sciences}
  \bibinfo{volume}{104}~(\bibinfo{number}{1}) (\bibinfo{year}{2007})
  \bibinfo{pages}{36--41}.

\bibitem[{Amiri et~al.(2013)Amiri, Hossain, Crawford, and
  Wigand}]{amiri2013community}
\bibinfo{author}{B.~Amiri}, \bibinfo{author}{L.~Hossain},
  \bibinfo{author}{J.~W. Crawford}, \bibinfo{author}{R.~T. Wigand},
  \bibinfo{title}{Community Detection in Complex Networks: Multi--objective
  Enhanced Firefly Algorithm}, \bibinfo{journal}{Knowledge-Based Systems}
  \bibinfo{volume}{46} (\bibinfo{year}{2013}) \bibinfo{pages}{1--11}.

\bibitem[{Lancichinetti et~al.(2009)Lancichinetti, Fortunato, and
  Kert{\'e}sz}]{lancichinetti2009detecting}
\bibinfo{author}{A.~Lancichinetti}, \bibinfo{author}{S.~Fortunato},
  \bibinfo{author}{J.~Kert{\'e}sz}, \bibinfo{title}{Detecting the overlapping
  and hierarchical community structure in complex networks},
  \bibinfo{journal}{New Journal of Physics}
  \bibinfo{volume}{11}~(\bibinfo{number}{3}) (\bibinfo{year}{2009})
  \bibinfo{pages}{033015}.

\bibitem[{Huang et~al.(2011)Huang, Sun, Han, and Feng}]{huang2011density}
\bibinfo{author}{J.~Huang}, \bibinfo{author}{H.~Sun}, \bibinfo{author}{J.~Han},
  \bibinfo{author}{B.~Feng}, \bibinfo{title}{Density-based shrinkage for
  revealing hierarchical and overlapping community structure in networks},
  \bibinfo{journal}{Physica A: Statistical Mechanics and its Applications}
  \bibinfo{volume}{390}~(\bibinfo{number}{11}) (\bibinfo{year}{2011})
  \bibinfo{pages}{2160--2171}.

\bibitem[{Yang et~al.(2013)Yang, Di, Liu, and Liu}]{yang2013hierarchical}
\bibinfo{author}{B.~Yang}, \bibinfo{author}{J.~Di}, \bibinfo{author}{J.~Liu},
  \bibinfo{author}{D.~Liu}, \bibinfo{title}{Hierarchical community detection
  with applications to real-world network analysis}, \bibinfo{journal}{Data \&
  Knowledge Engineering} \bibinfo{volume}{83} (\bibinfo{year}{2013})
  \bibinfo{pages}{20--38}.

\bibitem[{Kim and Wilhelm(2013)}]{kim2013spanning}
\bibinfo{author}{J.~Kim}, \bibinfo{author}{T.~Wilhelm},
  \bibinfo{title}{Spanning tree separation reveals community structure in
  networks}, \bibinfo{journal}{Physical Review E}
  \bibinfo{volume}{87}~(\bibinfo{number}{3}) (\bibinfo{year}{2013})
  \bibinfo{pages}{032816}.

\bibitem[{Zhang and Wang(2015)}]{zhang2014mining}
\bibinfo{author}{Z.~Zhang}, \bibinfo{author}{Z.~Wang}, \bibinfo{title}{Mining
  overlapping and hierarchical communities in complex networks},
  \bibinfo{journal}{Physica A: Statistical Mechanics and its Applications}
  \bibinfo{volume}{421} (\bibinfo{year}{2015}) \bibinfo{pages}{25--33}.

\bibitem[{Kim and Kim(2015)}]{kim2015detecting}
\bibinfo{author}{P.~Kim}, \bibinfo{author}{S.~Kim}, \bibinfo{title}{Detecting
  overlapping and hierarchical communities in complex network using
  interaction-based edge clustering}, \bibinfo{journal}{Physica A: Statistical
  Mechanics and its Applications} \bibinfo{volume}{417} (\bibinfo{year}{2015})
  \bibinfo{pages}{46--56}.

\bibitem[{Smyth and White(2005)}]{smyth2005spectral}
\bibinfo{author}{S.~Smyth}, \bibinfo{author}{S.~White}, \bibinfo{title}{A
  spectral clustering approach to finding communities in graphs}, in:
  \bibinfo{booktitle}{Proceedings of the 5th SIAM International Conference on
  Data Mining}, \bibinfo{pages}{76--84}, \bibinfo{year}{2005}.

\bibitem[{Hu et~al.(2008)Hu, Li, Zhang, Fan, and Di}]{hu2008community}
\bibinfo{author}{Y.~Hu}, \bibinfo{author}{M.~Li}, \bibinfo{author}{P.~Zhang},
  \bibinfo{author}{Y.~Fan}, \bibinfo{author}{Z.~Di}, \bibinfo{title}{Community
  detection by signaling on complex networks}, \bibinfo{journal}{Physical
  Review E} \bibinfo{volume}{78}~(\bibinfo{number}{1}) (\bibinfo{year}{2008})
  \bibinfo{pages}{016115}.

\bibitem[{Zhou et~al.(2014)Zhou, Martin, and Pan}]{zhou2014evidential}
\bibinfo{author}{K.~Zhou}, \bibinfo{author}{A.~Martin},
  \bibinfo{author}{Q.~Pan}, \bibinfo{title}{Evidential Communities for Complex
  Networks}, in: \bibinfo{booktitle}{Information Processing and Management of
  Uncertainty in Knowledge-Based Systems}, \bibinfo{organization}{Springer},
  \bibinfo{pages}{557--566}, \bibinfo{year}{2014}.

\bibitem[{Zhou et~al.(2015)Zhou, Martin, Pan, and Liu}]{zhou2015median}
\bibinfo{author}{K.~Zhou}, \bibinfo{author}{A.~Martin},
  \bibinfo{author}{Q.~Pan}, \bibinfo{author}{Z.-g. Liu}, \bibinfo{title}{Median
  evidential $c$-means algorithm and its application to community detection},
  \bibinfo{journal}{Knowledge-Based Systems} \bibinfo{volume}{74}
  (\bibinfo{year}{2015}) \bibinfo{pages}{69--88}.

\bibitem[{Liu et~al.(2009)Liu, Jiang, and Kot}]{liu2009multi}
\bibinfo{author}{M.~Liu}, \bibinfo{author}{X.~Jiang}, \bibinfo{author}{A.~C.
  Kot}, \bibinfo{title}{A multi-prototype clustering algorithm},
  \bibinfo{journal}{Pattern Recognition}
  \bibinfo{volume}{42}~(\bibinfo{number}{5}) (\bibinfo{year}{2009})
  \bibinfo{pages}{689--698}.

\bibitem[{Wang et~al.(2014)Wang, Chen, and Mei}]{wang2014incremental}
\bibinfo{author}{Y.~Wang}, \bibinfo{author}{L.~Chen}, \bibinfo{author}{J.-P.
  Mei}, \bibinfo{title}{Incremental Fuzzy Clustering With Multiple Medoids for
  Large Data}, \bibinfo{journal}{Fuzzy Systems, IEEE Transactions on}
  \bibinfo{volume}{22}~(\bibinfo{number}{6}) (\bibinfo{year}{2014})
  \bibinfo{pages}{1557--1568}.

\bibitem[{Gao et~al.(2013)Gao, Wei, Hu, Mahadevan, and Deng}]{gao2013modified}
\bibinfo{author}{C.~Gao}, \bibinfo{author}{D.~Wei}, \bibinfo{author}{Y.~Hu},
  \bibinfo{author}{S.~Mahadevan}, \bibinfo{author}{Y.~Deng}, \bibinfo{title}{A
  modified evidential methodology of identifying influential nodes in weighted
  networks}, \bibinfo{journal}{Physica A: Statistical Mechanics and its
  Applications} \bibinfo{volume}{392}~(\bibinfo{number}{21})
  (\bibinfo{year}{2013}) \bibinfo{pages}{5490--5500}.

\bibitem[{Zhou et~al.(2009)Zhou, L{\"u}, and Zhang}]{zhou2009predicting}
\bibinfo{author}{T.~Zhou}, \bibinfo{author}{L.~L{\"u}}, \bibinfo{author}{Y.-C.
  Zhang}, \bibinfo{title}{Predicting missing links via local information},
  \bibinfo{journal}{The European Physical Journal B}
  \bibinfo{volume}{71}~(\bibinfo{number}{4}) (\bibinfo{year}{2009})
  \bibinfo{pages}{623--630}.

\bibitem[{Pan et~al.(2010)Pan, Li, Liu, and Liang}]{pan2010detecting}
\bibinfo{author}{Y.~Pan}, \bibinfo{author}{D.-H. Li}, \bibinfo{author}{J.-G.
  Liu}, \bibinfo{author}{J.-Z. Liang}, \bibinfo{title}{Detecting community
  structure in complex networks via node similarity}, \bibinfo{journal}{Physica
  A: Statistical Mechanics and its Applications}
  \bibinfo{volume}{389}~(\bibinfo{number}{14}) (\bibinfo{year}{2010})
  \bibinfo{pages}{2849--2857}.

\bibitem[{Chen et~al.(2009)Chen, Za{\"\i}ane, and Goebel}]{chen2009detecting}
\bibinfo{author}{J.~Chen}, \bibinfo{author}{O.~R. Za{\"\i}ane},
  \bibinfo{author}{R.~Goebel}, \bibinfo{title}{Detecting Communities in Social
  Networks Using Max-Min Modularity.}, in: \bibinfo{booktitle}{SDM},
  vol.~\bibinfo{volume}{3}, \bibinfo{organization}{SIAM},
  \bibinfo{pages}{20--24}, \bibinfo{year}{2009}.

\bibitem[{Scripps et~al.(2007)Scripps, Tan, and
  Esfahanian}]{scripps2007exploration}
\bibinfo{author}{J.~Scripps}, \bibinfo{author}{P.-N. Tan},
  \bibinfo{author}{A.-H. Esfahanian}, \bibinfo{title}{Exploration of link
  structure and community-based node roles in network analysis}, in:
  \bibinfo{booktitle}{Data Mining, 2007. ICDM 2007. Seventh IEEE International
  Conference on}, \bibinfo{organization}{IEEE}, \bibinfo{pages}{649--654},
  \bibinfo{year}{2007}.

\bibitem[{Newman(2006{\natexlab{b}})}]{newman2006finding}
\bibinfo{author}{M.~E. Newman}, \bibinfo{title}{Finding community structure in
  networks using the eigenvectors of matrices}, \bibinfo{journal}{Physical
  review E} \bibinfo{volume}{74}~(\bibinfo{number}{3})
  (\bibinfo{year}{2006}{\natexlab{b}}) \bibinfo{pages}{036104}.

\bibitem[{Raghavan et~al.(2007)Raghavan, Albert, and Kumara}]{raghavan2007near}
\bibinfo{author}{U.~N. Raghavan}, \bibinfo{author}{R.~Albert},
  \bibinfo{author}{S.~Kumara}, \bibinfo{title}{Near linear time algorithm to
  detect community structures in large-scale networks},
  \bibinfo{journal}{Physical Review E}
  \bibinfo{volume}{76}~(\bibinfo{number}{3}) (\bibinfo{year}{2007})
  \bibinfo{pages}{036106}.

\bibitem[{Rosvall and Bergstrom(2008)}]{Rosvall29012008}
\bibinfo{author}{M.~Rosvall}, \bibinfo{author}{C.~T. Bergstrom},
  \bibinfo{title}{Maps of random walks on complex networks reveal community
  structure}, \bibinfo{journal}{Proceedings of the National Academy of
  Sciences} \bibinfo{volume}{105}~(\bibinfo{number}{4}) (\bibinfo{year}{2008})
  \bibinfo{pages}{1118--1123}.

\bibitem[{Krishnapuram and Keller(1993)}]{krishnapuram1993possibilistic}
\bibinfo{author}{R.~Krishnapuram}, \bibinfo{author}{J.~M. Keller},
  \bibinfo{title}{A possibilistic approach to clustering},
  \bibinfo{journal}{Fuzzy Systems, IEEE Transactions on}
  \bibinfo{volume}{1}~(\bibinfo{number}{2}) (\bibinfo{year}{1993})
  \bibinfo{pages}{98--110}.

\bibitem[{Pal et~al.(2005)Pal, Pal, Keller, and Bezdek}]{pal2005possibilistic}
\bibinfo{author}{N.~R. Pal}, \bibinfo{author}{K.~Pal}, \bibinfo{author}{J.~M.
  Keller}, \bibinfo{author}{J.~C. Bezdek}, \bibinfo{title}{A possibilistic
  fuzzy $c$-means clustering algorithm}, \bibinfo{journal}{Fuzzy Systems, IEEE
  Transactions on} \bibinfo{volume}{13}~(\bibinfo{number}{4})
  (\bibinfo{year}{2005}) \bibinfo{pages}{517--530}.

\bibitem[{Nepusz et~al.(2008)Nepusz, Petr{\'o}czi, N{\'e}gyessy, and
  Bazs{\'o}}]{nepusz2008fuzzy}
\bibinfo{author}{T.~Nepusz}, \bibinfo{author}{A.~Petr{\'o}czi},
  \bibinfo{author}{L.~N{\'e}gyessy}, \bibinfo{author}{F.~Bazs{\'o}},
  \bibinfo{title}{Fuzzy communities and the concept of bridgeness in complex
  networks}, \bibinfo{journal}{Physical Review E}
  \bibinfo{volume}{77}~(\bibinfo{number}{1}) (\bibinfo{year}{2008})
  \bibinfo{pages}{016107}.

\bibitem[{Fan et~al.(2007)Fan, Li, Zhang, Wu, and Di}]{fan2007accuracy}
\bibinfo{author}{Y.~Fan}, \bibinfo{author}{M.~Li}, \bibinfo{author}{P.~Zhang},
  \bibinfo{author}{J.~Wu}, \bibinfo{author}{Z.~Di}, \bibinfo{title}{Accuracy
  and precision of methods for community identification in weighted networks},
  \bibinfo{journal}{Physica A: Statistical Mechanics and its Applications}
  \bibinfo{volume}{377}~(\bibinfo{number}{1}) (\bibinfo{year}{2007})
  \bibinfo{pages}{363--372}.

\bibitem[{Lancichinetti et~al.(2008)Lancichinetti, Fortunato, and
  Radicchi}]{lancichinetti2008benchmark}
\bibinfo{author}{A.~Lancichinetti}, \bibinfo{author}{S.~Fortunato},
  \bibinfo{author}{F.~Radicchi}, \bibinfo{title}{Benchmark graphs for testing
  community detection algorithms}, \bibinfo{journal}{Physical Review E}
  \bibinfo{volume}{78}~(\bibinfo{number}{4}) (\bibinfo{year}{2008})
  \bibinfo{pages}{046110}.

\bibitem[{Brin and Page(1998)}]{brin1998anatomy}
\bibinfo{author}{S.~Brin}, \bibinfo{author}{L.~Page}, \bibinfo{title}{The
  anatomy of a large-scale hypertextual Web search engine},
  \bibinfo{journal}{Computer networks and ISDN systems}
  \bibinfo{volume}{30}~(\bibinfo{number}{1}) (\bibinfo{year}{1998})
  \bibinfo{pages}{107--117}.

\bibitem[{Zachary(1977)}]{zachary1977information}
\bibinfo{author}{W.~W. Zachary}, \bibinfo{title}{An information flow model for
  conflict and fission in small groups}, \bibinfo{journal}{Journal of
  anthropological research}  (\bibinfo{year}{1977}) \bibinfo{pages}{452--473}.

\end{thebibliography}

\end{document}